\documentclass[aps,pra,epsf,superscriptaddress,amsmath,amssymb,amsfonts,twocolumn,showpacs]{revtex4-1}
\usepackage{titlesec}
\usepackage{graphicx}
\usepackage{epsfig}
\usepackage{dcolumn}
\usepackage{bm}
\usepackage{braket}
\usepackage{amsmath}
\usepackage{amssymb}
\usepackage[titletoc]{appendix}

 \usepackage{multirow}

\def\be{\begin{equation}}
\def\ee{\end{equation}}

 \def\a{\alpha}
\begin{document}
\title{Quench-induced resonant tunneling mechanisms of bosons\\ in an optical lattice with harmonic confinement}

\author{G.M. Koutentakis}
\affiliation{Zentrum f\"{u}r Optische Quantentechnologien,
Universit\"{a}t Hamburg, Luruper Chaussee 149, 22761 Hamburg,
Germany}\affiliation{The Hamburg Centre for Ultrafast Imaging,
Universit\"{a}t Hamburg, Luruper Chaussee 149, 22761 Hamburg,
Germany}
\author{S.I. Mistakidis}
\affiliation{Zentrum f\"{u}r Optische Quantentechnologien,
Universit\"{a}t Hamburg, Luruper Chaussee 149, 22761 Hamburg,
Germany}
\author{P. Schmelcher}
\affiliation{Zentrum f\"{u}r Optische Quantentechnologien,
Universit\"{a}t Hamburg, Luruper Chaussee 149, 22761 Hamburg,
Germany} \affiliation{The Hamburg Centre for Ultrafast Imaging,
Universit\"{a}t Hamburg, Luruper Chaussee 149, 22761 Hamburg,
Germany}

\date{\today}

\begin{abstract}

The non-equilibrium dynamics of small boson ensembles in a one-dimensional optical lattice is explored  
upon a sudden quench of an additional harmonic trap from strong to weak confinement. 
We find that the competition between the initial localization and the repulsive interaction leads to a resonant response  
of the system for intermediate quench amplitudes, corresponding to avoided crossings in the many-body eigenspectrum with varying  
final trap frequency. In particular, we show that these avoided crossings can be utilized to prepare the system in a desired state.   
The dynamical response is shown to depend on both the interaction strength 
as well as the number of atoms manifesting the many-body nature of the tunneling dynamics.

\end{abstract}

 \maketitle


\section{Introduction}

Recent experimental advances in ultracold atomic gases have provided novel ways to examine the static properties and the 
non-equilibrium dynamics of correlated many-body systems \cite{Bloch,Polkovnikov,Weidemuller,Hung,Ronzheimer}.  
In particular, optical lattice potentials are a prominent feature of ultracold experiments, as they allow for the study of the correlated tunneling dynamics   
and its dependence on the interparticle interaction \cite{Ronzheimer,Meinert,Folling,Mistakidis,Mistakidis1}.  
Systems consisting of small ensembles of atoms offer the opportunity, both theoretically \cite{Mistakidis,Mistakidis1} and experimentally \cite{Zurn1,Zurn3}  
to identify and track microscopic (quantum) mechanisms due to their finite size and the absence of finite temperature effects.  
On the other hand, quantum quenches \cite{Cheneau,Natu,Chen,Haller,Mahmud,Campbell,Mistakidis,Mistakidis1} enable us to study the dependence of the   
dynamical response on the perturbation amplitude applied to an equilibrium system.  
Therefore, quenched finite systems in combination with the appropriate lattice geometries can lead to new  
quantum effects, especially when the translational invariance of the lattice is broken.

A well-studied model that breaks the translational invariance consists of a lattice potential with an imposed harmonic trap 
\cite{Ronzheimer,Rigol1,Batrouni,Kashurnikov,Kollath,Wessel,Tschischik}.  
Concerning the dynamics, the ballistic expansion rate of a bosonic Mott-insulator trapped in such a composite trap after a quench of the trap frequency to a lower value 
has been shown to depend on the interparticle interaction \cite{Ronzheimer}. 
Furthermore, it has been demonstrated \cite{Tschischik} that in the limit of low filling factors the dynamics is equivalent to harmonically trapped bosons with a lattice-dependent 
effective mass. Both of the above-mentioned effects emerge when the harmonic confinement is relatively weak compared to the interparticle repulsion. However, a   
so far largely unexplored theme is   
the competition between the harmonic confinement and the interaction strength, which favors different spatial configurations.  
An intriguing question would thereby be whether this competition can be exploited to obtain a high level of controllability of such a system  
and, as a consequence whether the out-of-equilibrium dynamics can be utilized to achieve specific state preparations. 

In the present work we consider a small ensemble of bosons confined in an optical lattice subjected to an additional strong harmonic confinement 
and investigate the dynamics induced by a quench from strong to weak confinement.  
We first analyze the many-body eigenspectrum for varying trap frequency, revealing the existence of narrow and wide avoided 
crossings between the many-body eigenstates. 
The dynamics of the interacting bosons shows distinct regions of weak and strong dynamical response. 
In particular, for increasing quench amplitude the system exhibits regions of a pronounced response in the vicinity of wide avoided crossings and 
sharper response peaks being a consequence of the corresponding narrow avoided crossings.  
Finally, it is shown that we can achieve specific state preparation by utilizing the narrow avoided crossings. Appropriately selecting the post quench 
trap frequency it is possible to couple the initial state to a desired final one, allowing for a low-frequency and efficient population transfer between the two eigenstates. 
Finally, the quench-induced many-body dynamics changes significantly with varying particle number and interparticle repulsion, as the positions and widths of the avoided crossings are shifted,  
giving rise to further variability and controllability of the dynamics.  
The results presented in this work are obtained by employing the Multi-Configuration Time-Dependent Hartree method for bosons (MCTDHB) \cite{Alon,Alon1}. 

The structure of the paper is as follows.
In section \ref{sec:theor_frame} we provide the underlying theoretical framework of our work. 
Section \ref{sec:4bs_3w} presents our triple well results, both for the static case and for the quench induced dynamics.
In section \ref{sec:seven_well} we present the generalization of our results for multi-well traps and, finally, in section \ref{sec:conclusions} we summarize and give an 
outlook. Appendix A describes our computational method.  

\section{Theoretical Framework} \label{sec:theor_frame}

In the present section we shall briefly discuss our theoretical framework. First, we introduce the many-body Hamiltonian (see \ref{sec:hamilt}) of our system. Then, the wavefunction 
representation in terms of a time-independent number state basis (\ref{subsec:NS_con}) is outlined. Finally, the basic observables (\ref{sec:rel_quantities}) used for the 
interpretation of the dynamics are explained. 

\subsection{The Hamiltonian} \label{sec:hamilt}

The many-body Hamiltonian of $N$ bosons trapped in a one-dimensional lattice potential with an imposed harmonic trap reads
\be
\begin{split}
H=&\sum_{i=1}^N \left( - \frac{\hbar^2}{2 m} \frac{\partial^2}{\partial x_i^2} + V_0 \sin^2 (k x_i) + \frac{m \omega^2}{2} x_i^2\right) \\
&+g \sum_{i=1}^N \sum_{j=i+1}^N \delta(x_i-x_j), \label{hamil}
\end{split}
\ee
where $x_i$ denotes the position of the $i$-th particle. The optical lattice potential is characterized by its depth $V_0$ and the corresponding wavenumber $k$. The imposed 
harmonic trap (parallel to lattice axis) depends on its frequency $\omega$ and confines the particles around the origin $x=0$. 
The effective $1D$ coupling strength of the contact interaction $g=\frac{2 \hbar^2 \a}{m a_\perp^2} \left(1-\frac{|\zeta (1/2)| \a}{\sqrt{2} a_\perp} \right)^{-1}$ \cite{Olshanii} 
can be manipulated via the transverse harmonic oscillator length ${a_ \bot } = \sqrt
{\frac{\hbar }{{M{\omega _ \bot }}}}$ (belonging to the strongly confined dimensions) \cite{Kim,Giannakeas} or by the $3D$ $s$-wave scattering length $\a$ via a Feshbach 
resonance \cite{Duine,Chin}.

To induce the dynamics we utilize the following scheme: The system is initially prepared in the ground state of the many-body Hamiltonian (see Eq.\ref{hamil}).  
Then, at $t=0$ we instantaneously change the trap frequency $\omega$ to a lower value and let the system    
evolve under the new Hamiltonian. 

Throughout this work we shall employ the recoil energy $E_\textrm{R}=\hbar^2 k^2/ (2 m)$, the inverse wavevector $k^{-1}$ and the bosonic mass $m$ as the units of the energy, 
length and mass, respectively. Hard wall boundary conditions are imposed at $x_\pm=\pm S \pi /2 k^{-1}$, where $S$ denotes the number of lattice sites. The depth of the 
lattice is fixed to $V_0=9 E_\textrm{R}$, thus including three localized single particle states.

\subsection{Number state expansion} \label{subsec:NS_con}

Using MCTDHB we calculate the many-body wavefunction $|\Psi (t) \rangle$ with respect
to a time-dependent basis consisting of variationally optimized single particle functions (SPFs), (for more information see Appendix A and  \cite{Alon,Alon1}).
However, for the analysis of our results it is preferable to project the numerically-obtained $|\Psi (t) \rangle$ in a time-independent number state basis
of single particle states localized on each lattice site. These localized states are constructed using the subset of     
delocalized eigenstates with $b$ nodes for each lattice site, i.e.  $| \psi_i^{(b)}\rangle_{g=0}$, $i \in \lbrace 1,\dots,S \rbrace$. 
In the absence of a harmonic confinement this subset of eigenstates belong to the $b$-th Bloch band of the system. To obtain a set of localized 
states we diagonalize the band-projected position operator 
$\hat{X}^{(b)} = \hat{P}^{(b)} \hat{x} \hat{P}^{(b)}$, where the operator $\hat{P}^{(b)} = \sum_{k = 1}^{S} \big| \psi_i^{(b)} \big\rangle_{g=0} \big\langle \psi_i^{(b)} \big|_{g=0}$ projects onto  
the band $b$ \cite{Wannier1,Wannier2,Wannier3}. In the following, we refer to the eigenstates $| \phi^{(b)}_s \rangle$, $s \in \lbrace 1,\dots,S \rbrace$  
of $\hat{X}^{(b)}$ as the single-band Wannier states of the deformed lattice. 
The corresponding $N$-body number state basis reads
\be
\begin{split}
\Big| \bigotimes_{b_1} n_1^{(b_1)}, &\dots, \bigotimes_{b_S} n_S^{(b_S)} \Big\rangle =\sum_{i=1}^{N!} \frac{ \hat{\mathcal{P}}_i \left( \bigotimes_{j=1}^N \big| \phi^{(b_j)}_{s_j} \big\rangle \right) }{ \sqrt{N! \prod_{b,s} n^{(b)}_s! \;}},
\end{split}
\ee
where the operator $\hat{\mathcal{P}}_i$ performs the $i$-th permutation of $N$ elements and $n^{(b)}_s$ refers to the number occupation of the Wannier state $\big| \phi^{(b)}_{s} \big\rangle$.
To simplify the notation we shall make the following assumptions. We omit the superscript if no Wannier state or only the Wannier states belonging to the $0$-th band are occupied,  
and decompose the occupation number as $n^{(b_1)} \otimes n^{(b_2)} \otimes \cdots$ if more than a single Wannier state localized in a certain well is occupied. 
For instance, $|1^{(0)}, 1^{(0)} \otimes 1^{(1)}, 1^{(2)} \rangle $ refers to the four-particle state of the triple-well where the ground states of the left and the middle well, the first excited state 
of the middle well, as well as the second excited state of the right well are each occupied by one boson. For later convenience we also denote by  
$| \vec{n} \rangle_S $ ($| \vec{n} \rangle_A $) the parity symmetric (antisymmetric) combination of the states 
$| \vec{n} \rangle = |\bigotimes n_1 , \bigotimes n_2, \dots, \bigotimes n_S \rangle$ and $| \bigotimes n_S , \bigotimes n_{S-1} \dots, \bigotimes n_1 \rangle$.

In the presence of a harmonic confinement, the Wannier number states are not uniquely ordered with respect to their energy expectation value, as argued in the following.    
Indeed, let us consider a system of four bosons in a triple-well. For  strong harmonic confinement and $g=0$ there are five number state subsets:    
$h_0=\{|0,4,0 \rangle\}$, $h_1=\{|1,3,0 \rangle_{S,A} \}$, $h_2=\{|2,2,0 \rangle_{S,A}, |1,2,1 \rangle \}$, $h_3=\{|3,1,0 \rangle_{S,A}, |2,1,1 \rangle_{S,A} \}$ 
and $h_4=\{|4,0,0 \rangle_{S,A}, |3,0,1 \rangle_{S,A}, |2,0,2 \rangle \}$ ordered in increasing energy. On the other hand, for an 
interacting gas with vanishing harmonic confinement there are four subsets of 
number states energetically ordered by the multiplicity of bosons that reside in each well: single pairs  
$i_{SP}=\{|2,1,1 \rangle, \circlearrowright\ \}$, double pairs $i_{DP}=\{|2,2,0 \rangle, \circlearrowright \} \}$, triplets $i_T=\{|3,1,0 \rangle, \circlearrowright\ \}$ 
and quadruplets $i_Q=\{|4,0,0 \rangle, \circlearrowright\ \}$, where $\circlearrowright$ stands for site permutations. Note that, for interaction energies of 
the order of the band gap, also higher band excitations must be considered. 

From the above example it becomes evident that a reordering of the number states in energy takes place as the system passes from the one limiting case to the other. For instance, the state 
$|0,4,0 \rangle$ belonging to the classes $h_0$ and $i_Q$, is the most favorable state for strong confinement but, at the same time, the most unfavorable for strong  
interactions. As we shall see, this corresponding reordering process is the main reason for the resonant dynamics in the quenched system.  

\subsection{Observables} \label{sec:rel_quantities}

Let us now briefly introduce a few basic observables to be utilized in the following analysis, which are based on the one-body density $\rho^{(1)}(x;t)$.  
To quantify the time-evolution of inter- and intrawell modes we use the average position of the bosons in a spatial region $D$, $D=\lbrace x \in (x_i,x_f) \rbrace$   
\be
\langle x \rangle_D(t) = \braket{ \Psi (t) | \hat{x}_D | \Psi (t) } = \frac{1}{N_D} \int_{x_i}^{x_f} dx~ x \rho^{(1)}(x;t),
\ee
where $N_D=\int_{x_i}^{x_f} dx~ \rho^{(1)}(x;t)$ is the particle number within $D$, 
$\hat{x}_D=\int_{x_i}^{x_f} dx~ x \hat \Psi^\dagger(x) \hat\Psi(x)$ refers to a one-body operator and $\hat\Psi(x)$ is the field operator. 
This quantity offers a measure for the collective displacement of the 
atoms (see ref. \cite{Kohn} for the dipole mode) within a prescribed region of the lattice.

Further, we introduce the position variance in a spatial region $D$ 
\be
\begin{split}
\sigma_{x,D}^2(t)&= \braket{ \Psi (t) | \hat{x}^2_D | \Psi (t) } - \braket{ \Psi (t) | \hat{x}_D | \Psi (t) }^2 \\
&=\frac{1}{N_D} \int_{x_i}^{x_f} dx~ x^2 \rho^{(1)}(x;t) - \langle x \rangle^2_D(t), \label{var}
\end{split}
\ee
which measures the expansion and contraction of the atomic cloud (see ref. \cite{Abraham2} for the breathing mode) within $D$.
Note the appearance of the one-body operator $\hat{x}^2_D=\int_{x_i}^{x_f} dx~ x^2 \hat \Psi^\dagger(x) \hat\Psi(x)$.  
The position variance over the whole lattice $\sigma_{x,L}^2(t)$ quantifies a 'global breathing' mode, composed of intersite tunneling  
and intrasite breathing and dipole modes, which contribute altogether to the total contraction and expansion of the atomic cloud.  
$\sigma_{x,L}^2(t)$ thus offers a measure for the net dynamical response of the system. 

To measure the impact of the quench we also define the time-averaged position 
variance 
\be
\begin{split}
\overline{\sigma_{x,L}^2} - \sigma_{x,L}^2(0) \equiv& \frac{1}{T}\int^T_0 dt ~\sigma_{x,L}^2(t) -\sigma_{x,L}^2(0) \\
\underset{T \to \infty}{=}& \sum_{i>j} \text{Re} \left( c_i^* ~ {}_{\omega}\langle \Psi_i | \hat{x}^2_L | \Psi_j \rangle_{\omega} ~ c_j \right),
\end{split}
\ee
which describes how far the system is in average from its initial state. Here, $\ket{\Psi_i}_\omega$ is the $i$-th excited stationary eigenstate of the post quench Hamiltonian
and $\ket{\Psi}=\sum_i c_i \ket{\Psi_i}_\omega$ holds. 
We have also used the parity symmetry which implies $\langle x \rangle_L(t)=\langle x \rangle_L(0)=0$ 
(since the initial state is the ground state of the Hamiltonian before the quench).  
Finally, the temporal variance of the position variance 
\be
\begin{split}
\Delta_T \lbrace \sigma_{x,L}^2 \rbrace \equiv &\frac{1}{T}\int^T_0 dt ~\left( \sigma_{x,L}^2(t) - \overline{\sigma_{x,L}^2} \right)^2 \\
\underset{T \to \infty}{=} &\frac{1}{2} \sum_{i>j} \left|  c_i^* ~ {}_{\omega}\langle \Psi_i | \hat{x}^2_L | \Psi_j \rangle_{\omega} ~ c_j  \right|^2
\end{split} \label{var_var}
\ee
quantifies how much the state fluctuates around its average configuration during the evolution.  
$\Delta_T \lbrace \sigma_{x,L}^2 \rbrace$ thus measures the intensity of the dynamical processes and of defect formations following a   
quench.
The theoretical limit $T \to \infty$ is replaced in practice by a finite evolution time $T$ where 
$\overline{\sigma_{x,L}^2}$ and $\Delta_T \lbrace \sigma_{x,L}^2 \rbrace$ have converged.

\section{Quench dynamics within a triple well} \label{sec:4bs_3w}

As a prototype system exhibiting characteristic quench induced dynamics, we use a  
system of four harmonically trapped bosons in a triple well potential. We first investigate the eigenspectra  
of the system with varying trap frequency (section \ref{subsec:gs_con}), which are subsequently related to the    
dynamics induced by the quench (section \ref{sec:quench_dynamics}). 

\subsection{Eigenspectra} \label{subsec:gs_con}

The eigenstate spectrum of the Hamiltonian of Eq. \ref{hamil} depends on both the trap frequency $\omega$ and the interaction strength $g$ (see also sec. \ref{subsec:NS_con}). 
To interpret the dynamics caused by a quench of the frequency $\omega$ of the imposed 
harmonic trap we analyze how the eigenstate spectrum depends on $\omega$. 
\begin{figure}[h]
\includegraphics[width=0.45\textwidth]{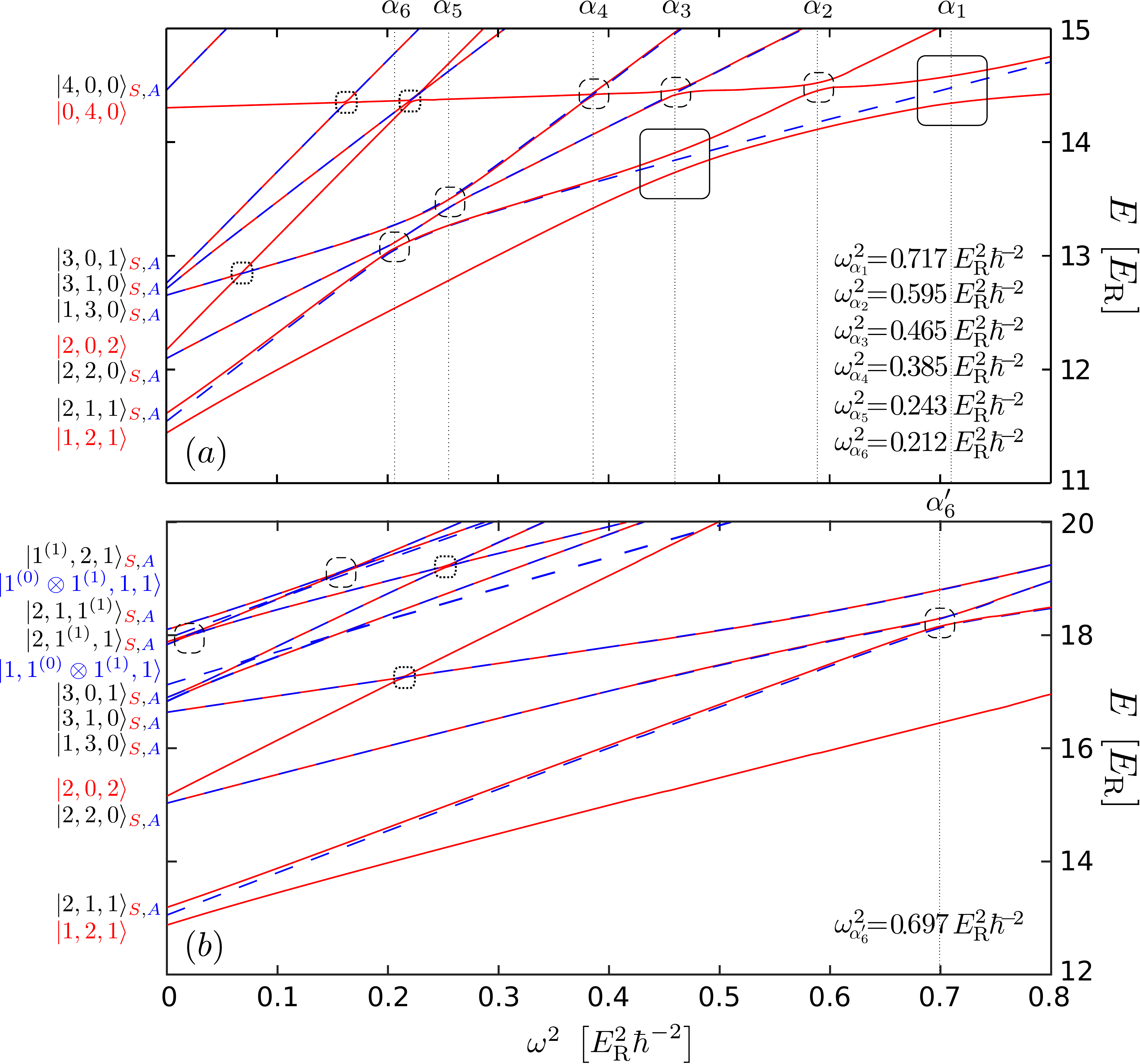}
\caption[]{(color online) Dependence of the eigenenergies of the Hamiltonian of Eq. \ref{hamil} on the trap frequency $\omega^2$ for a system consisting of $N=4$ bosons trapped in a triple well with additional harmonic confinement.  
The lowest ($a$) $15$ eigenenergies for $g=1 ~E_\textrm{R} k^{-1}$ and ($b$) $20$ eigenenergies for $g=4 ~E_\textrm{R} k^{-1}$ are shown. 
Solid (dashed) lines represent even (odd) parity eigenstates.  
Very narrow crossings (width smaller than $4 \times 10^{-3} E^2_\textrm{R} \hbar^{-2}$) are denoted by dotted boxes, narrow avoided crossings 
are indicated by dashed boxes and wide avoided crossings are denoted by solid boxes.  
$\alpha_i$, $i=1,...,6$ indicate the position of each avoided crossing (see legend).  
The dominant number state contribution of each eigenstate at $\omega=0$ is indicated on the left hand side of each figure. 
}
\label{fig:3w_frg}
\end{figure}

Fig. \ref{fig:3w_frg}($a$) shows the $\omega$ dependence of the eigenenergies for $g=1~E_\textrm{R} k^{-1}$. 
For later convenience we denote each of the even-parity eigenstates as $|\Psi_i \rangle$, where $i \in \{0,\dots,8 \}$ refers to their energetical order for a given $\omega$. 
In the following, we focus on the even-parity part of the spectrum $|\Psi_i \rangle$ (see Fig. \ref{fig:3w_frg}($a$)), which contains the ground state $|\Psi_0\rangle$.
At $\omega=0$ the number states within each class ($i_X$, $X=SP,DP,T,Q$) are very close in energy, their minor energetical difference being caused by the respective avoided crossings  
(and the boundary conditions of the 
triple well), and thus the eigenstates are a superposition of the corresponding Wannier number states.  
In particular, each of the eigenstates possesses a dominant contribution from a particular Wannier number state of class $i_X$ (presented on the left hand side of Fig. \ref{fig:3w_frg}($a$)). 
For $\omega > 0$, the eigenenergies increase linearly (proportional to the number of bosons in the side wells) with $\omega^2$. 
The population of the dominant contribution of each eigenstate increases with $\omega^2$ (at the expense of the contribution of other number states which belong to the same class) 
up to the point of complete dominance and therefore saturation.  
This behaviour with increasing $\omega^2$ occurs unless we encounter an avoided crossing with another eigenstate.  
Indeed, as it can be seen in Fig. \ref{fig:3w_frg}($a$) at such avoided crossings (being denoted by $C\in\{\alpha_i\}$, $i=1,...,6$) 
the two involved eigenstates exchange their character and therefore corresponding dominant Wannier number states. 
Indicative of this process is the fact that for $\omega^2>\omega^2_C$ the linear $\omega^2$-dependence of the eigenstates is restored and  
the corresponding slopes have been exchanged.
Moreover, the avoided crossings $\alpha_6$ and $\alpha_5$ due to their proximity exhibit a slightly different behaviour being referred to in the following as a composite avoided crossing 
$\{ \alpha_6 \alpha_5 \}$. 
We note that the avoided crossings which involve the ground state $| \Psi_0 \rangle$ are much wider (wide avoided crossings) compared to  
those involving excited states (narrow avoided crossings).
Despite the appearance of the above mentioned avoided crossings there are also very narrow avoided crossings   
possessing a corresponding width smaller than $4 \times 10^{-3} E^2_\textrm{R} \hbar^{-2}$.  

To interpret the eigenspectrum we employ the corresponding (three-site, lowest band) Bose-Hubbard model (BHM) \cite{BHM1,BHM2}   
\be 
\begin{split}
\hat{H}_{\text{BHM}}=&-J (\hat a^\dagger_1 \hat a_{2} + a^\dagger_2 \hat a_{3} +h.c.)\\
&+  \sum_{i=1}^3 \left[ \epsilon (i-2)^2 + e_0 \right] \hat n_i \\
&+ \frac{U}{2} \sum_{i=1}^3 \hat n_i( \hat n_i-1), \label{Bose_Hubbard}
\end{split}
\ee
where $\hat{a}_i$ ($\hat{a}^{\dagger}_i$) denotes the annihilation (creation) operator that annihilates (creates) a particle in the state $|\phi_i^{(0)} \rangle$ (e.g. $i=1$ refers to the leftmost well) 
and $\hat{n}_i=\hat{a}^\dagger_i \hat{a}_i$ corresponds to the particle number operator.
The Hubbard parameters $J$, $U$, $\epsilon$ and $e_0$ refer to the inter-site hopping, intra-site interaction, site offset energy and the zero point energy 
of the ensemble, respectively. 
We remark that all of the presented results are obtained within the MCTDHB framework
for the continuum space Hamiltonian of Eq. \ref{hamil} and we only refer to discrete
models (BHM, see Eq. \ref{Bose_Hubbard}), to interpret and compare our findings. 

The classification in terms of the interparticle interaction ($i_X$) provides information on the energy expectation value of the corresponding states 
for a vanishing harmonic confinement.  
Furthermore, the classification in terms of strong harmonic confinement provides information on how the energy of each Wannier number state depends on the trap frequency, i.e.   
$E\{ h_d \}(\omega)=E\{ h_d \}|_{\omega=0}+ d \epsilon$ ($\epsilon\propto\omega^2$).  
For instance, at $J=0$ all the crossings become exact (since all individual $n_i$ are conserved, see eq. \ref{Bose_Hubbard}) and we can evaluate the expected position 
for each one, e.g. the states $|1,2,1 \rangle$ and $|1,3,0\rangle_S$ are expected 
to cross at $\epsilon=4 U$.
For $J \neq 0$ these exact crossings become avoided and their widths can be approximated by the coupling between the involved states, e.g. 
$\langle 1, 2, 1 | \hat{H}_{\text{BHM}}|1,3,0\rangle_S = - \sqrt{3/2} J$, (see the wide avoided crossing in Fig. \ref{fig:3w_frg}($a$) at $\omega^2=\omega^2_{\alpha_3}$).  
Narrow avoided crossings emerge from higher order transitions yielding a nonlinear coupling in $J$, e.g. 
although $\langle 0, 4, 0 | \hat{H}_{\text{BHM}} |1, 2, 1\rangle = 0$   
these states are coupled by the higher order transition $|0, 4, 0\rangle \overset{J}{\rightleftharpoons} |1, 3, 0\rangle_S \overset{J}{\rightleftharpoons} |1, 2, 1\rangle$  
leading to a coupling $\propto J^2$. Here, the hopping $J\ll U$, and therefore the corresponding avoided crossing observed at $\omega^2=\omega^2_{\alpha_2}$ 
(see Fig. \ref{fig:3w_frg}($a$)) is much narrower. 
On the other hand, the composite avoided crossing $\{\alpha_6\alpha_5\}$ can be interpreted as follows: At $\omega^2_{\alpha_6}$ a comparatively narrow avoided crossing between  
the eigenstates $|\Psi_1 \rangle$ and $|\Psi_2 \rangle$ takes place (exchange between the Wannier number states $| 2,1,1 \rangle_S$ and $| 2,2,0 \rangle_S$). 
A wide avoided crossing involving the eigenstates $|\Psi_1 \rangle$ and $|\Psi_3 \rangle$ follows at $\omega^2_{\alpha_6} \leq \omega^2 \leq  \omega^2_{\alpha_5}$    
(exchange between the Wannier number states $| 1,3,0 \rangle_S$, $| 2,1,1 \rangle_S$). Finally, at $\omega^2_{\alpha_5}$ a second narrow avoided crossing takes place between the eigenstates 
$|\Psi_2 \rangle$ and $|\Psi_3 \rangle$ (exchange between the number states $| 2,1,1 \rangle_S$ and $| 2,2,0 \rangle_S$). 
This behaviour stems from the fact that all these three states are degenerate for $J=0$, $\epsilon=U$, while for finite $J$ the coupling 
between $| 2,1,1 \rangle_S$ and $| 1,3,0 \rangle_S$ can be neglected.
To connect our Hamiltonian (Eq. \ref{hamil}) with the employed BHM (see Eq. \ref{Bose_Hubbard}) we mention here that the studied case 
of $V_0=9~E_\textrm{R}$ and $g=1~E_\textrm{R} k^{-1}$ corresponds 
to the BHM Hamiltonian with parameters $U \sim 0.13~E_\textrm{R}$, $J \sim 1.5\times10^{-2}~E_\textrm{R}$, yielding a ratio $U/J \sim 8.5$.
The offset induced by the imposed harmonic oscillator potential corresponds to $\epsilon/\omega^2 \sim 0.55 ~\hbar^2/E_\textrm{R}$. 
Finally, it can be shown that the energy of the $|0,4,0 \rangle$ state depends on the trap frequency as $\sim 1+\frac{\omega^2}{4 V_0}$, related to the zero   
point energy $e_0$ (see Eq. \ref{Bose_Hubbard}). 

For larger interaction strength, the many-body states, as shown in Fig. \ref{fig:3w_frg}($b$), become energetically higher due to the higher interaction energy among the bosons. 
Note that a certain number of states with higher band excitations possess lower energies from states of the lowest band, e.g. the state $\ket{0,4,0}$ does not participate in the lowest twenty 
eigenstates of the Hamiltonian with $g=4~E_\textrm{R} k^{-1}$ (see Fig. \ref{fig:3w_frg}($b$)). 
Indeed, for $g=4~E_\textrm{R} k^{-1}$ the BHM can still be applied since:    
a) there is no coupling to higher band excitations 
via hopping terms of the form $-J^{b,b'}_{ij} a^{\dagger (b)}_i a^{(b')}_j=0$ due to the orthogonality of the 
single-particle Bloch states that belong to different bands, and b)   
in the interaction part of the Hamiltonian, intra-band on-site interaction terms due to $U^{b,b,b,b'}_{i,i,i,i} a^{\dagger (b)}_i a^{\dagger (b)}_i a^{(b)}_i a^{(b')}_i$, 
where $|b-b'|=2j+1$, $j\in \mathbb{N}$ vanish because of parity symmetry within the corresponding well \cite{note}.  
This manifests itself in Fig. \ref{fig:3w_frg}($b$) by the absence of avoided crossings between states with higher band excitations ($1$-st excited band)   
and states with all particles in the $0$-th band (see for instance the very narrow avoided crossing at $\omega^2=0.25~E^2_\textrm{R} \hbar^{-2}$ involving 
the states $\ket{\Psi_7}$ and $\ket{\Psi_8}$).
The energies of the latter possess a similar dependence on the frequency $\omega^{2}$ as for $g=1~E_\textrm{R} k^{-1}$ but the position in terms of $\omega^2$ of the avoided crossings 
between these states (see Fig. \ref{fig:3w_frg}($a$)) has increased approximately  
three times the original value for $g=1~E_\textrm{R} k^{-1}$.   
The dependence of the eigenenergies involving higher band excitations is linear in $\omega^2$ with the proportionality factor $\epsilon^{(1)}=\epsilon$, see for instance  
$\ket{\Psi_0}$ and $\ket{\Psi_8}$ within the frequency interval $\omega^2\in[0.1,0.25~E^2_\textrm{R} \hbar^{-2}]$.  
Furthermore, the states with higher band excitations show both narrow and wide avoided crossings, e.g see the narrow avoided crossing at $\omega^2=0.02~E^2_\textrm{R} \hbar^{-2}$ involving the states 
$\ket{\Psi_8}$ and $\ket{\Psi_9}$ and the wide avoided crossing at $\omega^2\in[0.02,0.16~E^2_\textrm{R} \hbar^{-2}]$ between the states $\ket{\Psi_8}$ and $\ket{\Psi_{10}}$. 
The narrow avoided crossings refer to intraband tunneling within the ground 
band while the wide avoided crossings are related to interband tunneling  
within the first excited band.   
Finally, let us also note that the narrow avoided crossings between (the parity symmetric) states with higher band excitations  
comprise a composite avoided crossing (see the dashed boxes at $\omega^2 \sim 0.02~E_\textrm{R}^2 \hbar^{-2}$ and $\omega^2 \sim 0.16~E_\textrm{R}^2 \hbar^{-2}$ in Fig. \ref{fig:3w_frg}($b$)) 
similar to the $\{ \alpha_6 \alpha_5 \}$ avoided crossing  
observed for the ground band (see Fig. \ref{fig:3w_frg}($a$)). 

Non-negligible widths of avoided crossings between states of the ground band and those having a contribution from the first excited band can in principle be achieved by shifting the center  
of the harmonic trap.   
This is equivalent to a lattice tilt \cite{Murillo} where the parity symmetry of the system is broken and states of different parity become coupled.
According to Fig. \ref{fig:3w_frg}($a$), ($b$) the states with different parity, e.g. $| \vec{n} \rangle_S$ and $| \vec{n} \rangle_A$ are near degenerate, except for the case of avoided crossings, and 
consequently we expect that the spectrum is slightly altered for the case of broken parity symmetry.

\subsection{Quench induced dynamics} \label{sec:quench_dynamics}

After having examined the basic properties of the eigenstate spectrum we proceed by investigating the many-body dynamics when the system is subjected to an abrupt 
quench of the trap frequency $\omega$ to lower values.  
\begin{figure*}[ht]
\includegraphics[width=0.85\textwidth]{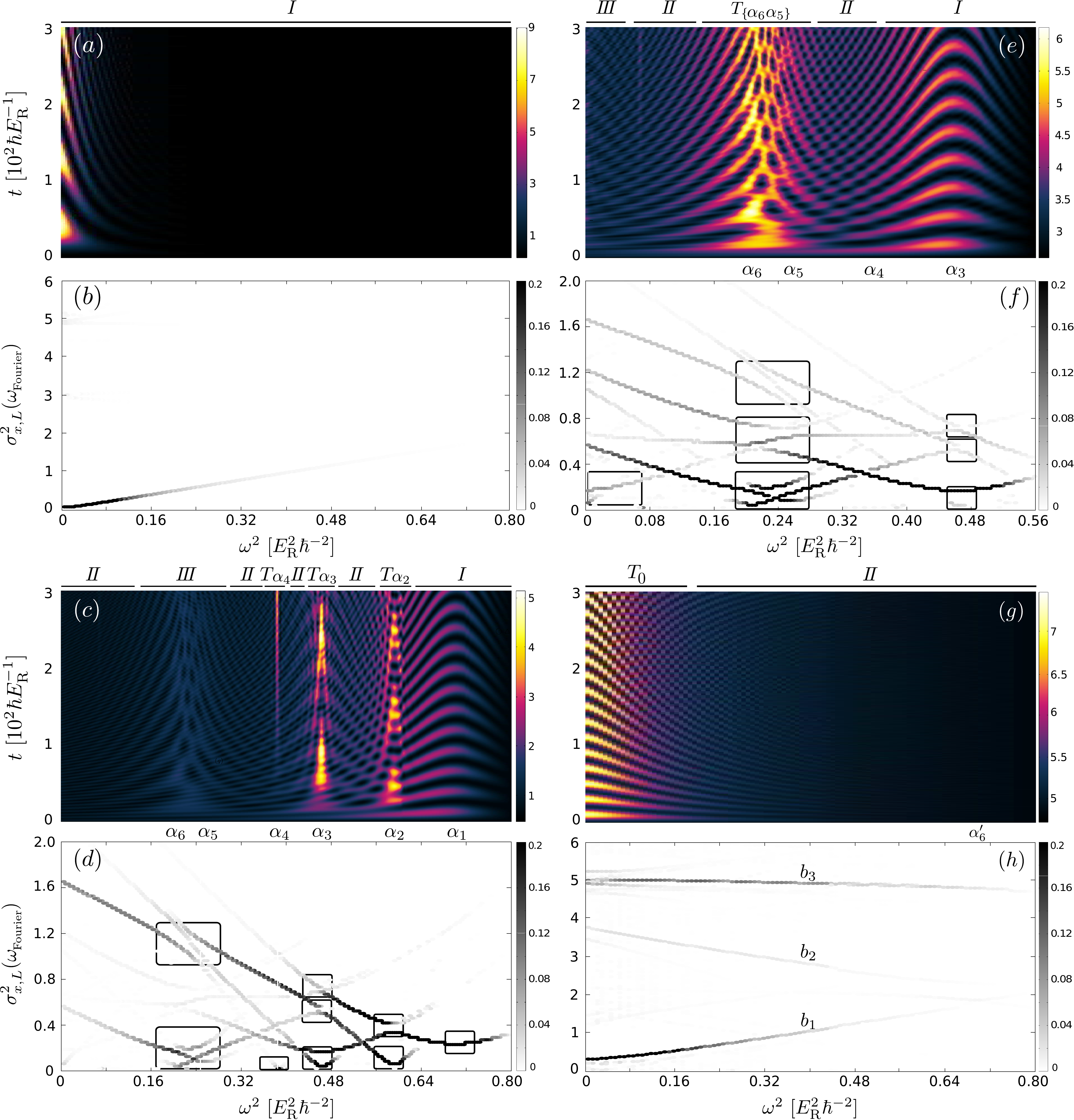}
\caption[]{(color online) ($a$,$c$,$e$,$g$) Time evolution of $\sigma_{x,L}^2(t)$ and ($b$,$d$,$f$,$h$) 
the corresponding spectra $\sigma_{x,L}^2(\omega_{\rm Fourier})$, as a function of the trap frequency $\omega^2$ after the quench.  
The system parameters used in each case correspond to ($a$,$b$) $g=0~ E_\textrm{R} k^{-1}$, $\omega_{i}^2=0.8 ~E_\textrm{R}^2 \hbar^{-2}$ (initial trap frequency), 
($c$,$d$) $g=1~ E_\textrm{R} k^{-1}$, $\omega_{i}^2=0.8 ~E_\textrm{R}^2 \hbar^{-2}$, 
($e$,$f$) $g=1~ E_\textrm{R} k^{-1}$, $\omega_{i}^2=0.58 ~E_\textrm{R}^2 \hbar^{-2}$ and ($g$,$h$) $g=4~ E_\textrm{R} k^{-1}$, $\omega_{i}^2=0.8 ~E_\textrm{R}^2 \hbar^{-2}$.  
$\alpha_i$, $i=1,...,6$ denote the positions of the corresponding avoided crossings (see also Fig. \ref{fig:3w_frg}). 
Note that on top of each of the ($a$,$c$,$e$,$h$) subfigures regions $I$, $II$, $III$, $T_C$ with $C\in\{\alpha_i\}$, are presented.  
In figures ($d$),($f$) the branches are organized by solid boxes denoted in Table I as segment $\alpha_{i,j}$, where the index $j=1,2,3$ introduces an energetically increasing order. }
\label{fig:3w_dyn_var}
\end{figure*}

For a strongly confined system initialized in the non-interacting ground state ($\ket\Psi \sim\prod|\phi_c^{(0)}\rangle$, $c$ stands for the middle well)    
the dynamical response of the system (for varying final trap frequency $\omega$) is shown in Fig. \ref{fig:3w_dyn_var}($a$) via $\sigma_{x,L}^2(t)$ (see Eq.(4)).   
This response can be described by the two single-particle states i.e. by   
$|\phi_c^{(0)}\rangle$ and the symmetric state $\frac{1}{\sqrt{2}} (|\phi_r^{(0)}\rangle + |\phi_\ell^{(0)}\rangle )$ ($l$, $r$ stand for the left and right wells respectively).   
For $\omega^2 > 0.08 ~E_\textrm{R}^2 \hbar^{-2} $ we have $\sigma_{x,L}^2(t)\approx\sigma_{x,L}^2(0)$   
showing that the system is unaffected by the quench and all the particles remain essentially localized in the center well ($\sim |\phi_c^{(0)}\rangle$). 
Only for quenches to very low trapping frequencies $\omega^2 < 0.08 ~E_\textrm{R}^2 \hbar^{-2}$, see in Fig. \ref{fig:3w_dyn_var}($a$) the particles diffuse to the outer wells  
and as a consequence the variance fluctuates significantly with time.  
The spectrum of the variance $\sigma_{x,L}^2(\omega_{\rm Fourier})$, presented in Fig. \ref{fig:3w_dyn_var}($b$), is dominated by a single frequency for all quenches, 
and this frequency corresponds to the Rabi frequency involving the energy difference between the two aforementioned states. 

Fig. \ref{fig:3w_dyn_var}($c$) presents $\sigma_{x,L}^2(t)$ as a function of $\omega^2$ for intermediate interactions $g=1 ~E_\textrm{R} k^{-1}$. 
Here, the ground state is dominated by the $|0,4,0 \rangle$ number state for the initial trap frequency $\omega^2_{i}=0.8$ (see Fig. \ref{fig:3w_frg}(a)).    
Regions of qualitatively different dynamical response with varying final trap frequency $\omega^2$ are manifest, denoted as $I$, $II$, $III$ and $T_C$ in Fig. \ref{fig:3w_dyn_var}($c$), 
thereby showing also a prominent difference from the $g=0$ case (see Fig. \ref{fig:3w_dyn_var}($a$)).  
Within the region denoted as type $I$ ($\omega^2> \omega^2_{\alpha_2}$) $\sigma_{x,L}^2(t)$ fluctuates prominently with time, indicating 
the presence of the global breathing mode.  
Here the fluctuations of $\sigma_{x,L}^2(t)$ are still characterized by a single frequency (see Fig. \ref{fig:3w_dyn_var}($d$)),  
and correspond to the Rabi oscillation region studied also in ref. \cite{Tschischik}. 
The regions of type $T_C$ (associated with a corresponding avoided crossing $C$) 
are characterized by a small frequency and high amplitude response during the evolution. 
Regions of type $II$ appear in between the regions of type $T_C$, e.g. $\omega_{\alpha_3}^2<\omega^2<\omega_{\alpha_2}^2$.
Here, the variance $\sigma_{x,L}^2(t)$ evolves with a multitude of frequencies while its amplitude is diminished compared to the case 
of regions $I$ and $T$.
Finally, within the region $II$ a region $III$ (linked to the composite avoided crossing $\{ \alpha_6 \alpha_5 \}$) of relatively strong response emerges.  

To gain more insight into the existence of the regions $II$, $III$ and $T_C$, we employ $\sigma_{x,L}^2(\omega_{\rm Fourier})$ shown in Fig. \ref{fig:3w_dyn_var}($d$). 
The connection of each of the branches observed in Fig. \ref{fig:3w_dyn_var}($d$) to the related eigenenergy differences is presented in detail in Table \ref{table:branches_g1}.  
We remark that only branches involving the state $|0,4,0 \rangle$ (being the dominant Wannier number state contribution of the initial state) 
are contributing significantly. 
The region $I$ is formed due to the avoided crossing $\alpha_1$ (see Fig. \ref{fig:3w_frg}(a)) and 
the bosons perform Rabi tunneling oscillations between the number states $|0,4,0 \rangle$ and $|1,3,0 \rangle_S$ (see segment $\alpha_{1,1}$ in Table $I$).  
Turning to the region $T_{\alpha_2}$ we observe that the dynamical response of the system  
is dominated by the low-frequency second-order tunneling process $| 1,2,1\rangle \rightleftharpoons |0,4,0 \rangle$ (see segment $\alpha_{2,1}$ in Table $I$). 
The formation of regions $II$ ($\omega^2_{\alpha_3} \leq \omega^2 \leq \omega^2_{\alpha_2}$) is caused by the fact that the tunneling modes 
$|1,2,1 \rangle \rightleftharpoons | 0,4,0\rangle$, 
$|1,3,0 \rangle_S \rightleftharpoons | 1,2,1\rangle$ and $|1,3,0 \rangle_S \rightleftharpoons | 0,4,0\rangle$ (see segments $\alpha_{2,1}$ and $\alpha_{2,2}$ in Table $I$) possess a similar amplitude 
(see also Fig. \ref{fig:3w_dyn_var} ($d$)) and the interference of these modes is destructive (on average) resulting in a weakened dynamical response.  
At $T_{\alpha_3}$ ($\omega^2=\omega^2_{\alpha_3}$) a low-frequency second order tunneling mode $| 2,2,0\rangle_S \rightleftharpoons |0,4,0 \rangle$ (see segment $\alpha_{3,1}$ in Table $I$) is observed.  
For larger quenches, i.e. $\omega^2 < \omega^2_{\alpha_3}$, the eigenstate dominated by $|0,4,0 \rangle$ does not couple with the remaining states of the 
eigenspectrum (see also Fig. \ref{fig:3w_frg}($a$)). 
Then, most of the processes appearing within this trap frequency regime    
are associated with the small $\ket{1,3,0}_S$ contribution to the initial state.    
The above give rise to the region $II$ which encompasses the $T_{\alpha_4}$ and $III$ regions.   
The $T_{\alpha_4}$ region is dominated by a low frequency mode being the third-order process 
$| 2,1,1\rangle_S \rightleftharpoons |0,4,0 \rangle$ (see segment $\alpha_{4,1}$).  
Note that even for very low detuning ($\omega^2\not=\omega^2_{\alpha_4}$) from the crossing $\alpha_4$,  
this third order process vanishes because:    
a) the width of the avoided crossing $\alpha_4$ is extremely narrow and b) the state $| 2,1,1\rangle_S$ has no contribution to the initial state. 
Finally, within the region $III$ that appears near the crossing $\{ \alpha_6 \alpha_5 \}$ (see Fig. \ref{fig:3w_dyn_var}($c$),($d$))    
the relevant low frequency tunneling processes are $| 2,1,1\rangle_S \rightleftharpoons |2,2,0 \rangle_S$
and $| 1,3,0\rangle_S \rightleftharpoons |2,2,0 \rangle_S$ (see segment $\{ \alpha_6 \alpha_5 \}_1$ in 
the appropriate regions and also Table \ref{table:branches_g1}).    
Here, the prominent dynamics of the state $|1,3,0 \rangle_S$ gives rise to a slightly  
increased response of the system in comparison to region $II$.  
\begingroup
\squeezetable
\begin{table*}
\begin{center}
\begin{ruledtabular}
\begin{tabular}{l l l}
\multicolumn{1}{c}{$\{ \alpha_6 \alpha_5 \}$} & \multicolumn{1}{c}{$\alpha_3$} & \multicolumn{1}{c}{$\alpha_2$} \\ 
\cline{1-1} \cline{2-2} \cline{3-3}\\
$	\{ \alpha_6 \alpha_5 \}_1 \begin{array}{l | l l}
		\multirow{2}{*}{$E_{21}$}  & | 1,3,0\rangle_S \rightleftharpoons |2,2,0 \rangle_S  & \omega^2 > \omega^2_{\alpha_5} \\
		                           & | 2,1,1\rangle_S \rightleftharpoons |2,2,0 \rangle_S  & \omega^2 < \omega^2_{\alpha_5} \vspace{1.5pt} \\
		\multirow{2}{*}{$E_{32}$}  & | 2,1,1\rangle_S \rightleftharpoons |2,2,0 \rangle_S  & \omega^2 > \omega^2_{\alpha_6} \\
		                           & | 1,3,0\rangle_S \rightleftharpoons |2,2,0 \rangle_S  & \omega^2 < \omega^2_{\alpha_6} \vspace{1.5pt} \\
		                   E_{31}  & | 1,3,0\rangle_S \rightleftharpoons |2,2,0 \rangle_S  & \omega^2 \in (\omega^2_{\alpha_6}, \omega^2_{\alpha_5} )
	\end{array} $ & 
$	\alpha_{3,1} \begin{array}{l l l}
		E_{32} & | 2,2,0\rangle_S \rightleftharpoons |0,4,0 \rangle & \hspace{5.0pt} \omega^2 \approx  \omega^2_{\alpha_3}\\
		E_{10} & | 1,3,0\rangle_S \rightleftharpoons |1,2,1 \rangle & \hspace{5.0pt} \omega^2 \approx  \omega^2_{\alpha_3}
	\end{array} $ & 	
$	\alpha_{2,1} \begin{array}{l l l}
		E_{21} & | 1,2,1\rangle \rightleftharpoons |0,4,0 \rangle & \hspace{5.0pt} \omega^2 \approx  \omega^2_{\alpha_2}
	\end{array} $ \vspace{2.5pt}  \\
$	\{ \alpha_6 \alpha_5 \}_2 \begin{array}{l | l l}
		\multirow{2}{*}{$E_{10}$} & |1,3,0 \rangle_S  \rightleftharpoons | 1,2,1\rangle & \hspace{5.0pt} \omega^2 > \omega^2_{\alpha_6} \\
		                          & |2,1,1 \rangle_S  \rightleftharpoons | 1,2,1\rangle & \hspace{5.0pt} \omega^2 < \omega^2_{\alpha_6} \vspace{1.5pt} \\ 
		\multirow{2}{*}{$E_{20}$} & |2,1,1 \rangle_S  \rightleftharpoons | 1,2,1\rangle & \hspace{5.0pt} \omega^2 \in (\omega^2_{\alpha_6}, \omega^2_{\alpha_5} ) \\
		                          & |2,2,0 \rangle_S  \rightleftharpoons | 1,2,1\rangle & \hspace{5.0pt} \omega^2 \notin (\omega^2_{\alpha_6}, \omega^2_{\alpha_5} ) \vspace{1.5pt} \\ 
		\multirow{2}{*}{$E_{30}$} & |2,1,1 \rangle_S  \rightleftharpoons | 1,2,1\rangle & \hspace{5.0pt} \omega^2 > \omega^2_{\alpha_5} \\
	                                  & |1,3,0 \rangle_S  \rightleftharpoons | 1,2,1\rangle & \hspace{5.0pt} \omega^2 < \omega^2_{\alpha_5} \\
	\end{array} $ & 
$	\alpha_{3,2} \begin{array}{l | l l}
		\multirow{2}{*}{$E_{21}$} & | 2,2,0\rangle_S \rightleftharpoons |1,3,0 \rangle_S              & \omega^2 < \omega^2_{\alpha_3} \\
		                          & | 0,4,0\rangle \hspace{5.0pt} \rightleftharpoons |1,2,1 \rangle   & \omega^2 > \omega^2_{\alpha_3} \vspace{1.5pt} \\
		E_{31}                    & | 1,3,0\rangle_S \rightleftharpoons |0,4,0 \rangle                & \omega^2 < \omega^2_{\alpha_3}
	\end{array} $ &
$	\alpha_{2,2} \begin{array}{l | l l}
		\multirow{2}{*}{$E_{10}$} & | 1,3,0\rangle_S \rightleftharpoons |0,4,0 \rangle & \omega^2 > \omega^2_{\alpha_2} \\
		                          & | 1,3,0\rangle_S \rightleftharpoons |1,2,1 \rangle & \omega^2 < \omega^2_{\alpha_2} \vspace{1.5pt} \\
		\multirow{2}{*}{$E_{20}$} & | 1,3,0\rangle_S \rightleftharpoons |1,2,1 \rangle & \omega^2 > \omega^2_{\alpha_2} \\
		                          & | 1,3,0\rangle_S \rightleftharpoons |0,4,0 \rangle & \omega^2 < \omega^2_{\alpha_2}
	\end{array} $ \vspace{2.5pt} \\ 
$	\{ \alpha_6 \alpha_5 \}_3 \begin{array}{l l l}
                E_{43} & | 1,3,0 \rangle_S \rightleftharpoons |0,4,0 \rangle & \hspace{5.0pt} \omega^2 < \omega^2_{\alpha_5}\\
		E_{42} & | 1,3,0 \rangle_S \rightleftharpoons |0,4,0 \rangle & \hspace{5.0pt} \omega^2 \in (\omega^2_{\alpha_6}, \omega^2_{\alpha_5} ) \\
		E_{41} & | 1,3,0 \rangle_S \rightleftharpoons |0,4,0 \rangle & \hspace{5.0pt} \omega^2 > \omega^2_{\alpha_6}
	\end{array} $ & 
$	\alpha_{3,3} \begin{array}{l | l l}
		\multirow{2}{*}{$E_{20}$} & | 1,3,0\rangle_S \rightleftharpoons |0,4,0 \rangle              & \hspace{5.0pt} \omega^2 > \omega^2_{\alpha_3} \\
		                          & | 2,2,0\rangle_S \rightleftharpoons |1,2,1 \rangle              & \hspace{5.0pt} \omega^2 < \omega^2_{\alpha_3} \vspace{1.5pt}\\
	        E_{30}                    & | 1,2,1\rangle \hspace{5.0pt} \rightleftharpoons |0,4,0 \rangle & \hspace{5.0pt} \omega^2 < \omega^2_{\alpha_3}
	\end{array} $ & \vspace{2.5pt} \\ \hline 
\multicolumn{1}{c}{$\alpha_4$} & \multicolumn{1}{c}{$\alpha_1$} & \multicolumn{1}{c}{$0$} \\
\cline{1-1} \cline{2-2} \cline{3-3}\\
\hspace{6.5pt} $\alpha_{4,1}$ \hspace{3.5pt} $E_{43}$ \hspace{0.5pt} $| 2,1,1\rangle_S \rightleftharpoons |0,4,0 \rangle$  \hspace{6.0pt} $\omega^2 \approx  \omega^2_{\alpha_4}$ &
\hspace{-2.5pt} $\alpha_{1,1}$ \hspace{-2.5pt} $E_{10}$ \hspace{1.0pt} $| 1,3,0 \rangle_S \rightleftharpoons |0,4,0 \rangle$ \hspace{6.0pt} $\omega^2 \approx  \omega^2_{\alpha_1}$ & 
\hspace{-2.5pt} $0_{1,1} \begin{array}{l l l}
		  E_{10} & | 2,1,1\rangle_S \rightleftharpoons |1,2,1 \rangle & \hspace{1.0pt} \omega \approx 0\\
		  E_{23} & | 2,2,0\rangle_S \rightleftharpoons |2,0,2 \rangle & \hspace{1.0pt} \omega \approx 0
	\end{array} $\\
\end{tabular}
\end{ruledtabular}
\end{center}
\caption[]{Energy differences and the corresponding main tunneling processes for each branch that appear in the spectra of Fig. \ref{fig:3w_dyn_var}($d$,$f$). 
Each avoided crossing is denoted by $\alpha_i, i=1,...,6$, while $0$ refers to the avoided crossings located at $\omega=0$.  
The branches are categorized into segments being denoted by a second subscript index $j$, i.e. $\alpha_{i,j}, j=1,2,3$.
The energy difference of a given branch is denoted by $E_{km} \equiv E_k -E_m$ 
followed by the corresponding main tunneling processes and the respective frequency domains.}
\label{table:branches_g1}
\end{table*}
\endgroup

To examine the case of the composite avoided crossing $\{ \alpha_6 \alpha_5 \}$ we initialize the system to the ground state  
$|\Psi_0 \rangle$ ($g=1~E_\textrm{R} k^{-1}$ and $\omega_{i}^2=0.560~E^2_\textrm{R} \hbar^{-2}$) being dominated by the Wannier number state $|1,3,0\rangle_S$. 
Fig. \ref{fig:3w_dyn_var}($e$) shows  
$\sigma^2_{x,L}(t)$ for different final trapping frequencies $\omega$. 
For $\omega^2>\omega^2_{\alpha_4}$ the system undergoes Rabi oscillations of varying amplitude (region $I$ in Fig. \ref{fig:3w_dyn_var}($e$)). As the quench amplitude  
increases the system transits smoothly via the region $II$ where the amplitude of the Rabi oscillations decreases to the region $T_{\{ \alpha_6 \alpha_5 \}}$   
where the response of the system is prominent.  
Note here that the region $T_{\{ \alpha_6 \alpha_5 \}}$ is broader than the previously mentioned regions of type $T_C$.  
Finally, for $\omega^2 < \omega^2_{\alpha_6}$ (region $II$) the response of the system becomes multi-mode and the fluctuations of 
$\sigma^2_{x,L}(t)$ are diminished. 

To identify the corresponding microscopic processes, Fig. \ref{fig:3w_dyn_var}($f$) presents $\sigma_{x,L}^2(\omega_{\rm Fourier})$.   
The region $I$ is dominated by the tunneling mode $| 1,3,0 \rangle_S \rightleftharpoons |1,2,1 \rangle$ (see segment $\alpha_{3,1}$) and it is related to the wide avoided crossing $\alpha_3$ 
(see solid boxes in Fig. \ref{fig:3w_frg}($a$)).  
As the quench amplitude increases the region $II$ appears for similar reasons as in the case of $\omega_i=0.8$ and $\omega^2\in(\omega_{\alpha_3},\omega_{\alpha_2})$. 
Within the region $T_{\{ \alpha_6 \alpha_5 \}}$ the behaviour of the system can be summarized as follows.
For $\omega^2 > \omega^2_{\alpha_5}$, the process $|1,3,0 \rangle_S \rightleftharpoons | 2,2,0\rangle_S$ dominates the dynamics, 
while at $\omega^2 \approx \omega^2_{\alpha_5}$ the additional mode
$| 2,1,1\rangle_S \rightleftharpoons | 2,2,0 \rangle_S$ appears.  
Next, at $\omega^2_{\alpha_6} \leq \omega^2 \leq \omega^2_{\alpha_5}$ the latter mode possesses two distinct frequencies, while  the process
$|1,3,0 \rangle_S \rightleftharpoons | 2,2,0\rangle_S$ possesses a higher frequency than in the case of $\omega^2 > \omega^2_{\alpha_5}$.  
Finally, at $\omega^2 \approx \omega^2_{\alpha_5}$ the low-amplitude mode $| 2,1,1\rangle_S \rightleftharpoons | 2,2,0 \rangle_S$ corresponds
to a single frequency and the process $|1,3,0 \rangle_S \rightleftharpoons | 2,2,0\rangle_S$
dominates the dynamics. 
Concluding, the response of the system at $T_{\{ \alpha_6 \alpha_5 \}}$ depends strongly on the post-quench trap frequency. 
For $\omega^2 < \omega^2_{\alpha_6}$ a region $II$ appears similarly to the case of $\omega_i=0.8$ and $\omega^2<\omega_{\alpha_3}$.   
For $\omega \approx 0$ two additional branches of low-frequency appear which correspond to 
the avoided crossings between $|1,2,1\rangle$ and $|2,1,1\rangle_S$ and between $|2,2,0\rangle$  and 
$|2,0,2\rangle_S$ (see segment $0_{1,1}$). These modes do not involve the major contribution to the initial 
state $|1,3,0\rangle_S$ and consequently possess very small amplitude (region $III$).

For higher interactions (here $g=4 ~E_\textrm{R} k^{-1}$), the dynamical response of the system, shown in Fig. \ref{fig:3w_dyn_var}($g$) via $\sigma_{x,L}^2(t)$, after 
a quench on $\omega$ ($\omega^2_i=0.8~ E_\textrm{R}^2 \hbar^{-2}$) shows a 
qualitatively different behaviour from the case of intermediate interactions. 
The initial state is dominated by $|1,2,1 \rangle$ and possesses significant contributions from higher band excitations, e.g.   
$|1,1^{(0)}\otimes 1^{(2)},1 \rangle$ and $|1, 2^{(1)},1 \rangle$ \cite{note1}. 
The system is essentially unperturbed for $\omega^2 > 0.16~ E_\textrm{R}^2 \hbar^{-2}$ and the evolution is characterized by multiple frequencies (see region $II$). 
Remarkably enough even for small quench amplitudes regions of Rabi oscillations (see Fig. \ref{fig:3w_dyn_var}($a$),($c$),($e$)) are absent.   
Only for quenches to $\omega \approx 0$, a prominent response is observed (region $T_0$ in Fig. \ref{fig:3w_dyn_var}($g$)). 
Fig. \ref{fig:3w_dyn_var}($h$) presents $\sigma_{x,L}^2(\omega_{\rm Fourier})$ where we observe (in contrast to the case of intermediate interactions) the appearance of only a few branches. 
The most dominant branch (denoted as $b_1$) refers to the tunneling mode $|1,2,1\rangle \rightleftharpoons |2,1,1 \rangle_S$ while  
the second branch (denoted as $b_2$) corresponds to the process $|1,2,1\rangle \rightleftharpoons |1,3,0 \rangle_S$.
These tunneling modes appear due to the avoided crossing at $\omega=0$ (see Fig. \ref{fig:3w_frg}($b$)).  
Finally, the third branch (denoted as $b_3$) corresponds to the dipole mode 
$|1,2,1\rangle \rightleftharpoons |1^{(1)},2,1 \rangle_S$. 
This dipole mode is induced by the minor shift of the side well caused by the quench and it is of single particle nature \cite{note2}.   
We remark here that in the case of strong interactions the tunneling dynamics is suppressed  
allowing the dipole mode to possess a prominent role in the course of the evolution (see branch $b_3$). 
\begin{figure*}[ht]
\includegraphics[width=0.7\textwidth]{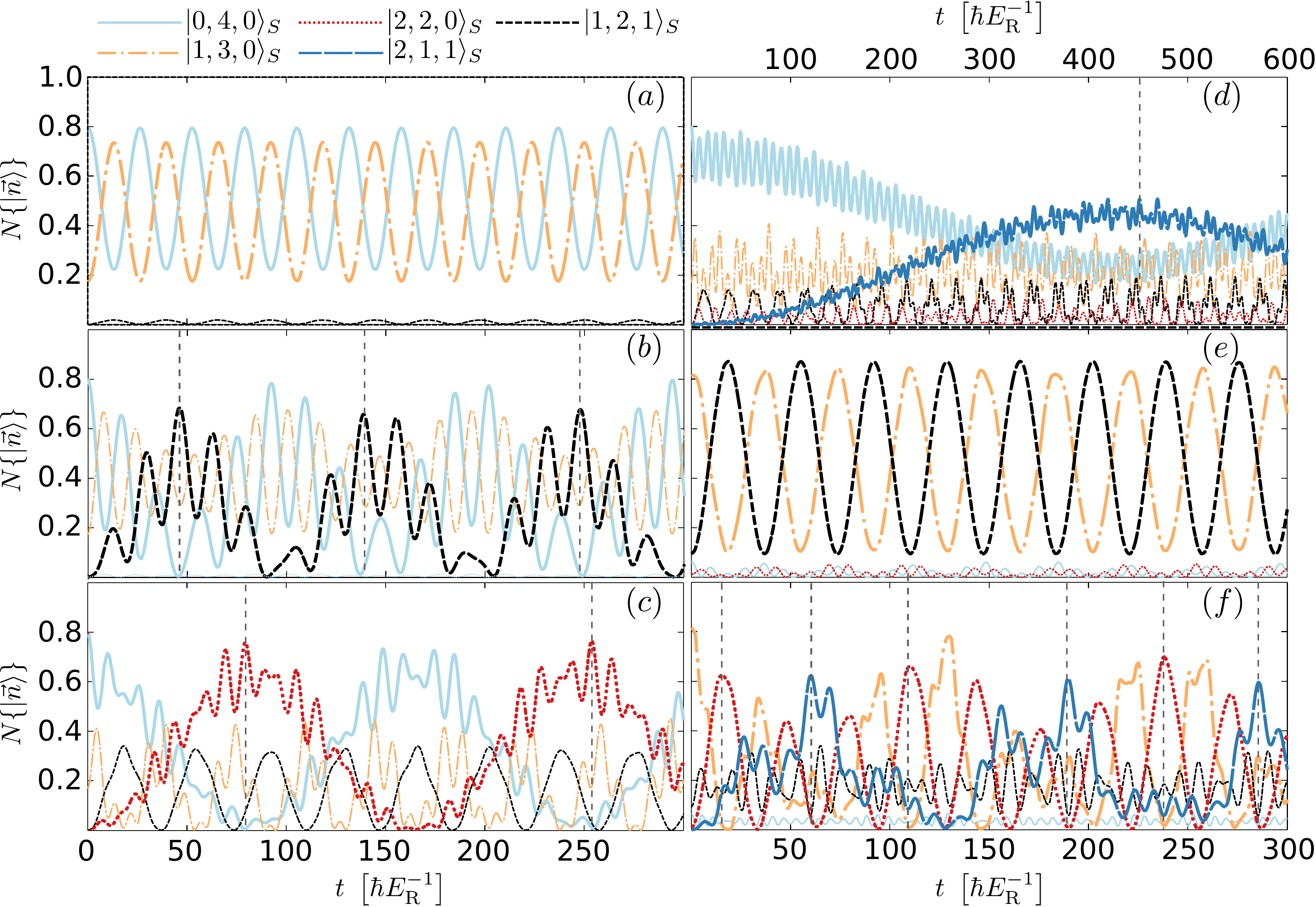}
\caption[]{(color online) Evolution of the dominant number state populations (see the legend above ($a$)) for $N=4$ interacting bosons ($g=1~ E_\textrm{R} k^{-1}$) confined in a triple well 
with additional harmonic confinement ($\omega^2_{i}=0.8~ E_\textrm{R}^2 \hbar^{-2}$). The dynamics is induced via a quench of the trap frequency  
to ($a$) $\omega^2=0.745~ E_\textrm{R}^2 \hbar^{-2}$ (region $I$), ($b$) $\omega^2=0.58~ E_\textrm{R}^2 \hbar^{-2}$ (region $T_{\alpha_2}$), ($c$) $\omega^2=0.48~ E_\textrm{R}^2 \hbar^{-2}$ (region $T_{\alpha_3}$) and   
($d$) $\omega^2=0.385~ E_\textrm{R}^2 \hbar^{-2}$ (region $T_{\alpha_4}$). Note that the scaling of the time axis in ($d$) is different from the other cases and appears on top of the figure.
The evolution of the number state populations is also shown for a quench from $\omega^2_{i}=0.56~ E_\textrm{R}^2 \hbar^{-2}$ to ($e$) $\omega^2=0.48~ E_\textrm{R}^2 \hbar^{-2}$ (region $I$) 
and ($f$) $\omega^2=0.265~ E_\textrm{R}^2 \hbar^{-2}$ (region $T_{ \{\alpha_6 \alpha_5 \}}$). For the identification of the different regions see Figure 2.}
\label{fig:3w_dyn_ns}
\end{figure*}

A natural next step is to investigate whether abrupt quenches can be used for state preparation.  
To achieve this goal let us examine the occupation of specific number states, i.e. $N \lbrace | \vec{n} \rangle \rbrace(t)=\left| \langle \vec{n} | \Psi(t) \rangle \right|^2$ during the 
evolution within the above-mentioned dynamical regions (see also Fig. \ref{fig:3w_dyn_var}).  
Fig. \ref{fig:3w_dyn_ns}($a$),($b$),($c$),($d$) show the dynamics of a system initialized in the ground state ($\omega^2_{i}=0.8~E^2_\textrm{R} \hbar^{-2}$, $g= 1~ E_\textrm{R} k^{-1}$) dominated by  
the number state $|0,4,0 \rangle$ while in Fig. \ref{fig:3w_dyn_ns}($e$),($f$) we consider the case 
($\omega^2_{i}=0.56~E^2_\textrm{R} \hbar^{-2}$, $g=1~ E_\textrm{R} k^{-1}$) in which the $|1,3,0 \rangle_S$  
possesses the dominant contribution to the ground state.
Fig. \ref{fig:3w_dyn_ns}($a$) shows $N \lbrace | \vec{n} \rangle \rbrace(t)$ for a quench within the region $I$ and near the avoided crossing $\alpha_1$. 
A population transfer from the number state $|0,4,0 \rangle$ to $|1,3,0 \rangle_S$ (see segment $\alpha_{1,1}$ in Table $I$) following Rabi oscillations is observed. 
Additional contributions stemming mainly from the state $|2,2,0\rangle_S$ are negligible.
On the other hand, Fig. \ref{fig:3w_dyn_ns}($b$) presents $N \lbrace | \vec{n} \rangle \rbrace(t)$ for a quench within the region $T_{\alpha_2}$. 
The dominant tunneling process corresponds to $|0,4,0 \rangle \rightleftharpoons|1,2,1\rangle$ (see segment $\alpha_{2,1}$), while the additional high frequency tunneling  
modes $|1,3,0 \rangle_S \rightleftharpoons |0,4,0 \rangle$, $|1,3,0 \rangle_S \rightleftharpoons |1,2,1\rangle$ coexist. As a consequence 
$|1,3,0 \rangle_S$ possesses a significant occupation during the dynamics. 
Fig. \ref{fig:3w_dyn_ns}($c$) shows $N \lbrace | \vec{n} \rangle \rbrace(t)$ for a quench within the region $T_{\alpha_3}$.
The main population transfer takes place between the number states $|0,4,0 \rangle$ and $|2,2,0\rangle_S$ (see segment $\alpha_{3,1}$). The influence 
of additional contributions to the initial state is smaller because:   
a) the frequency of the main tunneling mode $|0,4,0 \rangle \rightleftharpoons |2,2,0 \rangle_S$ is much lower than the frequency of the tunneling modes that couple 
the dominant state $\ket{0,4,0}$ with $\ket{1,3,0}_S$ and $\ket{1,2,1}$ and b) the tunneling mode $|1,3,0 \rangle_S \rightleftharpoons |1,2,1 \rangle$ (see segment $\alpha_{3,1}$) is 
pronounced due to the wide avoided crossing $\alpha_3$. 
Fig. \ref{fig:3w_dyn_ns}($d$) presents $N \lbrace | \vec{n} \rangle \rbrace(t)$ for a quench within the region $T_{\alpha_4}$.
We observe that the main population transfer takes place between the number states $|0,4,0 \rangle$ and $|2,1,1\rangle_S$. 
The frequency of the corresponding tunneling mode (see segment $\alpha_{4,1}$) is much lower compared to the other cases shown
in Fig. \ref{fig:3w_dyn_ns} and also compared to the remaining tunneling processes, e.g. $|1,3,0 \rangle_S \rightleftharpoons |2,2,0\rangle_S$, that appear in Fig. \ref{fig:3w_dyn_ns}($d$). 
Fig. \ref{fig:3w_dyn_ns}($e$) shows $N \lbrace | \vec{n} \rangle \rbrace(t)$ for a system initialized at $\omega_{i}^2=0.56~E^2_\textrm{R} \hbar^{-2}$ and following a quench within  
the region $T_{\alpha_3}$.
In this case, Rabi oscillations between the number states $|1,3,0 \rangle_S$ and $|1,2,1 \rangle$ (see segment $\alpha_{3,1}$) are observed. 
Finally, Fig. \ref{fig:3w_dyn_ns}($f$) illustrates $N \lbrace | \vec{n} \rangle \rbrace(t)$ (same initial state) for a quench within the 
region $T_{\{ \alpha_6 \alpha_5 \} }$.
Here, there are three dominant states, namely $|1,3,0 \rangle_S$, $|2,2,0\rangle_S$ and $|2,1,1\rangle_S$, which are coupled via the tunneling 
modes $|1,3,0 \rangle_S \rightleftharpoons |2,2,0\rangle_S$  
and $|2,2,0 \rangle_S \rightleftharpoons |2,1,1\rangle_S$ (see segment $\{ \alpha_6 \alpha_5 \}_1$). 
The total state of the system $\ket{\Psi(t)}$ is dominated within different time intervals by each of the above-mentioned 
number states.   
Concluding from the above, we note that it is possible to employ an abrupt quench of the trap frequency and achieve state preparation to one of the 
number states $| 1,3,0 \rangle_S$, $| 1,2,1 \rangle$, $| 2,2,0 \rangle_S$ and $| 2,1,1 \rangle_S$, with an adequate population $|\langle \vec{n} | \Psi(t) \rangle|^2 \geq 0.6$ 
(see the vertical dashed lines in Fig. \ref{fig:3w_dyn_ns}) by choosing properly the total evolution time.  
This implies that multimode evolution can be used for state preparation, as long as, the frequency of the desired 
transition is much lower than the competing population transfer processes during the dynamics. 
\begin{figure}[t]
\includegraphics[width=0.45\textwidth]{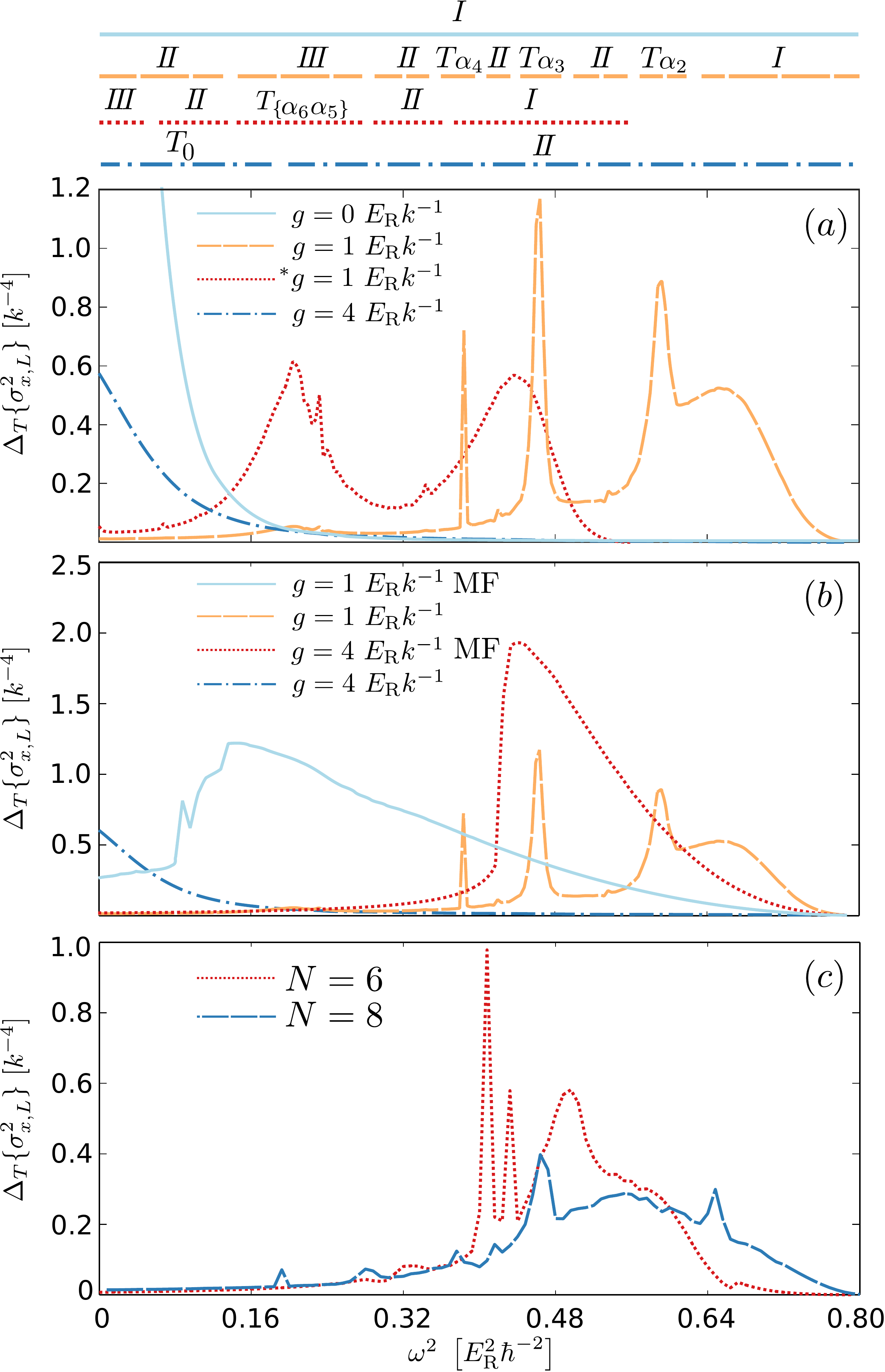}
\caption[]{(color online) $\Delta_T \lbrace \sigma_{x,L} \rbrace$ as a function of the post quench frequency $\omega^2$.
	Comparison between ($a$) different interaction strengths $g$ (see legend), 
	($b$) the mean field (MF) and MCTDHB results (see legend), and
	($c$) different particle numbers $N=6,8$ (see legend) for $g=0.5~ E_\textrm{R} k^{-1}$. 
	The symbol ${}^*$ denotes that the system is initialized in the ground state for    
	$\omega_{i,*}^2=0.56~ E_\textrm{R}^2 \hbar^{-2}$ instead of $\omega_{i}^2=0.8~ E_\textrm{R}^2 \hbar^{-2}$ being used otherwise.}
\label{fig:3w_dyn_res}
\end{figure}

To examine the dependence of the intensity of the dynamical processes on the quench amplitude, we employ   
$\Delta_T \lbrace \sigma_{x,L}^2 \rbrace$ (see Eq. \ref{var_var}) being a measure of the time-averaged dynamical response which depends solely on the parameters of the system.   
Fig. \ref{fig:3w_dyn_res} presents $\Delta_T \lbrace \sigma_{x,L}^2 \rbrace$ as a function of the post quench trap frequency $\omega^2$   
for different values of the relevant physical parameters. 

Fig. \ref{fig:3w_dyn_res}($a$) exhibits $\Delta_T \lbrace \sigma_{x,L}^2 \rbrace$ with varying quench amplitude for different interaction strengths $g$.    
In the non-interacting case (see also Fig. \ref{fig:3w_dyn_var}($a$)) the mean response of the system after a quench of the trap frequency is close to zero for a wide range of final 
trapping frequencies ($\omega^2>0.08~E^2_\textrm{R} \hbar^{-2}$). However, for large quench amplitudes ($\omega^2<0.08~E^2_\textrm{R} \hbar^{-2}$) the mean response increases strongly as the state 
$|\phi^{(0)}_c \rangle$ couples to the state $(|\phi^{(0)}_\ell \rangle + |\phi^{(0)}_r \rangle )/\sqrt{2}$.   
For the case of intermediate interactions ($g=1~E_\textrm{R} k^{-1}$) and initial trap frequency $\omega^2_{i}=0.8~E^2_\textrm{R} \hbar^{-2}$ (see also Fig. \ref{fig:3w_dyn_var}($c$)) 
we observe, as expected, that the different regions of dynamical response (i.e. $I$, $II$, $III$, $T_C$) yield a distinct averaged response.  
In particular, for quenches within the region $I$ the mean response of the system increases until it reaches its maximum value at $\omega^2_{\alpha_1}$.  
For $\omega^2 < \omega^2_{\alpha_1}$ the mean response of the system decreases.
The $T_{\alpha_2}$ region appears as a peak in $\Delta_T \lbrace \sigma_{x,L}^2 \rbrace$ because the interwell tunneling mode   
$|0,4,0\rangle  \rightleftharpoons |1,2,1\rangle$ becomes resonant.  
For $\omega^2_{\alpha_3}<\omega^2<\omega^2_{\alpha_2}$ the first region of type $II$ appears and $\Delta_T \lbrace \sigma_{x,L}^2 \rbrace$ exhibits 
a local minimum. As $\omega^2 \sim \omega^2_{\alpha_3}$ the mean response increases (at the $T_{\alpha_3}$ region)   
exhibiting a narrow peak followed by a region $II$ where the response is minimized. 
This process is repeated for $\omega^2 \sim \omega^2_{\alpha_4}$ within the region $T_{\alpha_4}$. The response peak at $T_{\alpha_4}$ is much narrower 
compared to the peaks in regions $T_{\alpha_3}$ and $T_{\alpha_2}$ 
since the resonant tunneling process $|0,4,0 \rangle \rightleftharpoons |2,1,1\rangle_S$ is of third order. 
Furthermore, the mean response of the system  
remains small for $\omega^2 < \omega^2_{\alpha_4}$, with the only remarkable feature being the slightly increased response observed in region $III$. 

For $g=1~E_\textrm{R} k^{-1}$ and $\omega^2_{i}=0.56~E^2_\textrm{R} \hbar^{-2}$ (see also Fig. \ref{fig:3w_dyn_var}($e$)) 
a different overall dynamical response is observed. 
The region of type $I$ exhibits a similar behaviour as for $\omega^2_{i}=0.8~E^2_\textrm{R} \hbar^{-2}$ but   
in this case the region $II$ follows the region $I$ for decreasing $\omega^2$ without the appearance of a $T$ region. 
For $\omega^2 \approx \omega^2_{\alpha_5}$, an increase in the mean response is observed as the system approaches the $T_{\{ \alpha_6 \alpha_5 \}}$ region. 
In contrast to the other $T_C$ regions the $T_{\{ \alpha_6 \alpha_5 \}}$ covers a wide  
range of trapping frequencies $\omega^2$ and exhibits a response peak at $\omega^2 = \omega^2_{\alpha_5}$ due to the tunneling 
mode $| 1, 3,0 \rangle_S \rightleftharpoons | 2,2,0 \rangle_S$ (see wide part of the crossing ${\{ \alpha_6 \alpha_5 \}}$).  
Finally, for larger quench amplitudes the response is minimized and increases slightly only at $\omega=0$ within the region $III$.
For high interactions ($g=4~E_\textrm{R} k^{-1}$) and initial trap frequency $\omega^2_{i}=0.8~E^2_\textrm{R} \hbar^{-2}$ the dynamical response of the system is 
completely different from the case of intermediate interactions.  
The initial state of the system is dominated by $|1,2,1 \rangle$ (which is also  
the main contribution to the ground state for $\omega^2 < \omega^2_{i}$).  
The system remains essentially unperturbed except for quenches in the proximity of $\omega \approx 0$ where 
an increasing response is observed due to the existence of the tunneling mode $|1,2,1 \rangle \rightleftharpoons |2,1,1 \rangle_S$ (see $T_0$ region).

Fig. \ref{fig:3w_dyn_res}($b$) presents a comparison of $\Delta_T \lbrace \sigma_{x,L}^2 \rbrace$ obtained within the mean-field approximation and  
the corresponding MCTDHB result i.e. taking into account the correlations.   
As it is clearly visible the two dynamical responses differ significantly.  
In particular, for intermediate interactions ($g=1$), the mean-field response 
of the system possesses a higher amplitude in comparison to the correlated case and continues to increase also beyond the regions $T_C$. 
Only for $\omega^2<\omega^2_{\alpha_3}$ it finally decreases almost abruptly to a small but  
finite value (which is still larger than the corresponding MCTDHB value).   
A similar behaviour is observed for higher interaction strengths. Here, the mean field 
approach overestimates even more the dynamical response of the system as a consequence of the increased interparticle repulsion.  
We conclude that the multiple resonant behaviour obtained within the correlated approach can not be captured by the mean-field 
approximation and consequently correlations between particles are crucial for the dynamics. 

Obviously, the behaviour of the system depends strongly on the particle number, since for different values of the latter the eigenstates and spectrum change overall.     
Fig. \ref{fig:3w_dyn_res}($c$) presents $\Delta_T \lbrace \sigma^2_{x,L} \rbrace$ 
for the particle numbers $N=6,8$ and $g=0.5~ E_\textrm{R} k^{-1}$.  
Both cases show a qualitatively similar response to the $N=4$ case presented in Fig. \ref{fig:3w_dyn_res}($a$). 
Indeed, high response peaks appear also here, indicating the existence of avoided crossings 
in the corresponding many-body spectrum. 
The overall response (see e.g. the response peaks) is altered with the number of particles, manifesting the many-body nature of the 
induced dynamics.  

To verify the applicability of our results for larger systems, in the following section, we shall consider multi-well setups  
consisting of a finite optical lattice and additional harmonic confinement. Then, we shall demonstrate that the character of the diffusion dynamics induced by a quench of the frequency 
of the imposed harmonic trap shows similar characteristics to the triple well case.  
\begin{figure}[h]
\includegraphics[width=0.48\textwidth]{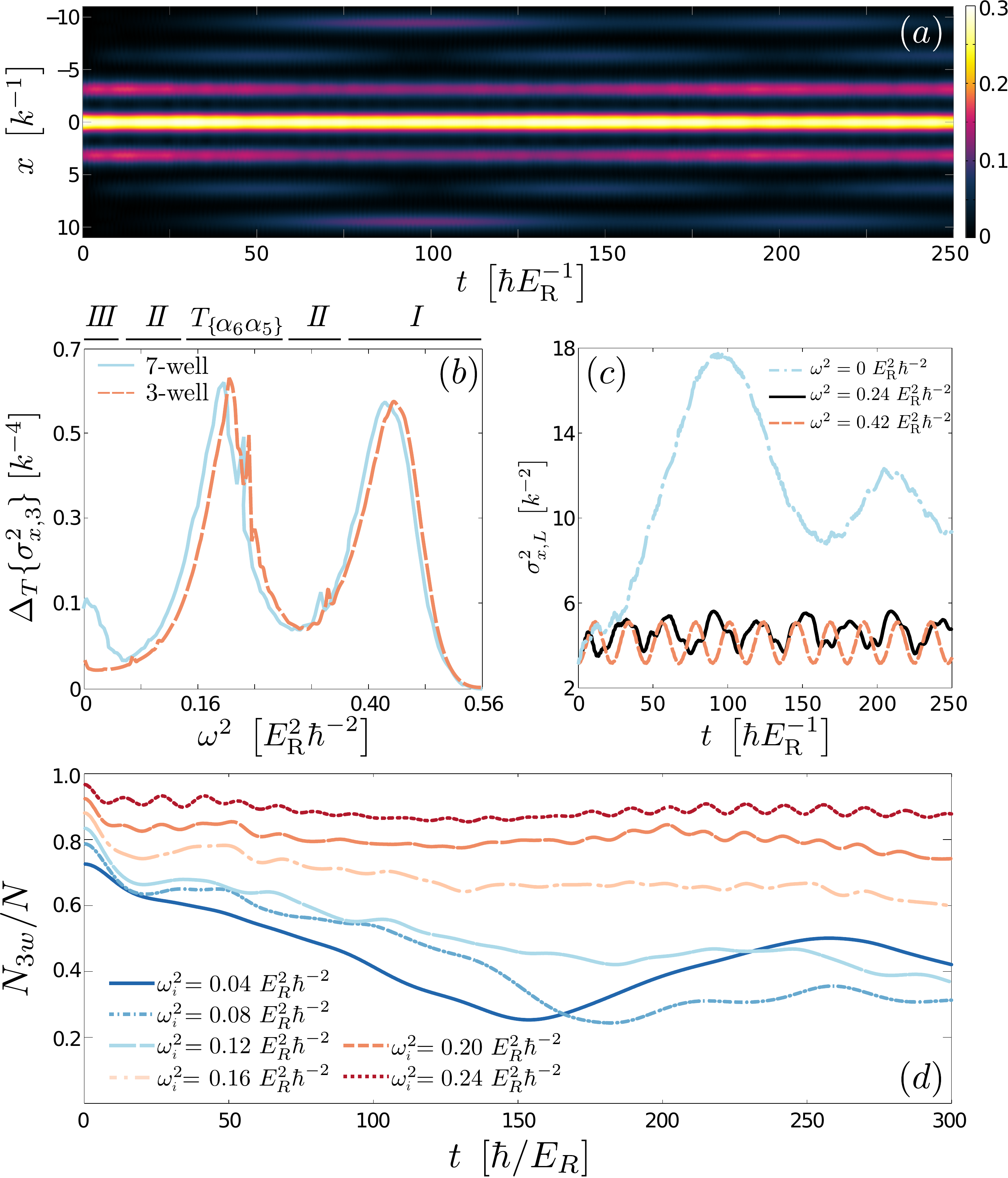}
\caption[]{(color online) ($a$) $\sqrt{\rho^{(1)}(x;t)}$ for $N=4$ interacting bosons ($g=1~E_\textrm{R} k^{-1}$) confined in a seven-well lattice with additional harmonic confinement.  
				The initial trap frequency is $\omega^2_{i}=0.56~E_\textrm{R}^2 \hbar^{-2}$ and we quench to $\omega=0$. 
				($b$) $\Delta_T \lbrace \sigma_{x,3w}^2 \rbrace$ with respect to the post quench $\omega^2$ for the triple well and the seven well setup 
				(see legend) with the same parameter 
				values, i.e. $\omega^2_{i}=0.56~E_\textrm{R}^2 \hbar^{-2}$ and $g=1~E_\textrm{R} k^{-1}$.
				($c$) $\sigma_{x,L}^2(t)$ for varying final trap frequency (see legend) for the seven well setup. ($d$) Evolution of the density fraction within 
				the triple-well region ($N_{3w}/N$) for a fifteen well with an imposed harmonic trap. The system is initialized in the ground state 
				with $N=5$, $g=1~E_\textrm{R} k^{-1}$ for varying initial trap frequency, $\omega_{i}^2$ (see legend), and the dynamics is induced  
				by a quench to $\omega^2=0.016~E_\textrm{R}^2 \hbar^{-2}$.}
\label{fig:7w_dyn_res}
\end{figure}

\section{Dynamics in multi-well traps} \label{sec:seven_well}

Let us consider four bosons confined in a seven-well  
lattice with additional harmonic confinement. The system is initially prepared in the ground state for  
$g=1~E_\textrm{R} k^{-1}$ and $\omega^2_{i}=0.56~E^2_\textrm{R} k^{-2}$, where all the particles are mainly localized in the  
center of the trap. The initial state is dominated by Wannier number states of the form $| 0,0,1, 3 ,0,0,0 \rangle_S$. 
Fig. \ref{fig:7w_dyn_res}(a) shows the dynamics of the ensemble after a quench of the trap frequency to zero.   
We observe that even in this extreme case the atoms remain essentially localized within 
the region of the triple well.   
In particular, a small portion of the particle density tunnels away from the central well and reaches the edge of the triple well region  
(where the boson gas is mainly localized) at $t\sim100~\hbar E_\textrm{R}^{-1}$ and subsequently 
expands ballistically beyond the three core sites (see Fig. \ref{fig:7w_dyn_res}($a$)).

To explore the mean dynamical response of the system Fig. \ref{fig:7w_dyn_res}($b$) illustrates $\Delta_T \lbrace \sigma^2_{x,3} \rbrace$ 
(where $3$ stands for the integration over the three core sites) for varying final trap frequency. To perform a direct 
comparison with the case of the triple well we also show the mean response of the latter with equal system parameters.  
A similar behaviour between the two cases is observed. However, we observe two deviations.  
Firstly, the corresponding response peaks are slightly shifted to a lower value of $\omega^2$. This effect can be 
explained by the artificial energy offset that the hard-wall boundary conditions introduce to the side wells with respect to the middle well. 
Indeed, this energy offset is reduced in the seven well case. 
Secondly, an increase of $\Delta_T \lbrace \sigma_{x,3}^2 \rbrace$ for $\omega \to 0$ is observed (see region $II$). 
This is due to the portion of atoms that expand beyond the triple well region (the dominant tunneling 
mode corresponds to $| 0,0,2,1,1,0,0 \rangle_S \rightleftharpoons | 0,1,1,1,1,0,0 \rangle_S$). 
Indeed, as can be seen from Fig. \ref{fig:7w_dyn_res}($a$) the unbound atomic density expands, then scatters at the hard-wall boundaries 
and finally re-enters the triple well.
This process enhances the amplitude of $\sigma_{x,3}^2$ and as a consequence the corresponding temporal variance. 

To investigate the dynamical response Fig. \ref{fig:7w_dyn_res}($c$) presents $\sigma_{x,L}^2(t)$ for different 
final trap frequencies. It is shown that for a quench to $\omega^2=0$ the portion of the density which is unbound 
to the triple well region expands ballistically. Indeed, $\sigma_{x,L}^2(t)$ after some critical value 
($\sigma_{x,L}^2(t_0=100)\approx 5 ~k^{-2}$) becomes linear for a finite time interval ($100<t<300~\hbar E_\textrm{R}^{-1}$). Furthermore, for quenches within the region of a response peak 
(e.g. see region $T_{\alpha_5}$ at $\omega^2\sim 0.25~E_\textrm{R}^2 \hbar^{-2}$ in Fig. \ref{fig:3w_frg}($a$)) or Rabi oscillations 
(e.g. see region $I$ at $\omega^2\sim 0.40 E_\textrm{R}^2 \hbar^{-2}$) the amplitude of $\sigma_{x,L}^2(t)$ is smaller than 
the critical one, thus showing the absence of a ballistically expanding fragment.

To further characterize the expansion dynamics we consider $N=5$ bosons confined in a fifteen-well lattice potential with an imposed  
harmonic trap. 
The dynamics, shown in Fig. \ref{fig:7w_dyn_res}($d$), is induced by a quench of the trap frequency and in particular we study quenches that refer to the same final trap  
frequency ($\omega^2=0.016~E_\textrm{R}^2 \hbar^{-2}$) but a different initial one. To compare the expansion dynamics we measure the fraction of the particle density within the three core sites during the 
dynamics, i.e. $N_{3w}(t)/N=\frac{1}{N}\int_{-3 \pi/2}^{3 \pi/2} dx \rho^{(1)}(x;t)$. Fig. \ref{fig:7w_dyn_res}($d$) shows that the above fragment of 
the particle density gets suppressed with increasing initial trap frequency. 
This indicates that an initially strongly confined bosonic ensemble remains after a quench of the 
trap frequency confined near the sites it was initially trapped into, which 
is a manifestation of the well-known self-trapping effect \cite{Smerzi,Raghavan,Albiez}. 

\section{Conclusions and outlook} \label{sec:conclusions}

We have investigated the eigenspectra and in particular the out-of-equilibrium quantum dynamics for a small ensemble of bosons confined in a 
lattice potential with an imposed harmonic trap. We hereby focus on the case of a strong harmonic confinement where the eigenstates become 
well-separated and are dominated by a single Wannier number state. 

In the non-interacting case, a significant tunneling dynamics is observed only for the case of a small final harmonic trapping.   
For intermediate interactions multiple avoided crossings with varying $\omega$ appear 
in the eigenspectrum which can be exploited to reveal a rich dynamics after quenching the trap frequency.
For relatively small quench amplitudes we observe Rabi oscillations caused by the wide avoided crossings between the ground and the first excited states.  
However, by using intermediate quench amplitudes we can utilize  
narrow avoided crossings involving solely excited states to selectively couple the initial state to a desired final state. The induced dynamics is characterized 
by multiple frequencies one of which is particularly slow and can be used to drive the system to a desired final state.  
For large quench amplitudes a multi-mode and low amplitude dynamical response is realized. 
In this case the number state with the dominant contribution to the initial state is an eigenstate of the final system (low amplitude dynamical response), while 
the remaining contributions to the initial state give rise to the observed multimode dynamics.
The deterministic preparation of the system in a desired Wannier number state is hindered by the fact that more tunneling 
modes are induced from additional contributions to the initial state. 
The case of stronger interparticle interactions with admixtures of a single excitation to the first excited band  
that do not couple in the eigenstate spectrum have been explored. The  
avoided crossings appear at higher trap frequencies and are narrower.   
The dynamics is different from the case of weak interactions, with higher band effects being more prominent and interwell tunneling being suppressed.

Let us comment on possible experimental implementations of our setup. In a corresponding ultracold gas experiment    
strongly interacting bosons are trapped in a   
one-dimensional superlattice.  
This superlattice can be formed by two retroreflected laser beams from which the first one possesses a large wavenumber and intensity 
(forming each supercell) compared to the second (forming each cell of the supercell). The above mentioned wavenumbers should be commensurate.
In this way, the potential landscape near the center of each supercell is similar to 
the one considered in the present study.    
Such a system may be implemented either by the use of holographic masks \cite{qmicroscope2} or by the 
modulation of the wavenumber, e.g. using accordion lattices \cite{Tli}. 
The trap frequency and the barrier height can be manipulated independently via the intensity of the lattice beams,   
and, finally, the interparticle interaction can be modulated via a magnetic Feshbach resonance.  
The corresponding static and dynamical properties of this state can then be measured with the recently developed single-site resolved imaging 
techniques \cite{qmicroscope1a,qmicroscope2,qmicroscope2a,qmicroscope3}. 
A experimental alternative would be to prepare $2N$ fermionic $^6$Li atoms in a microtrap \cite{Zurn1}, condense them into $N$ Li$_2$ bosonic Feshbach molecules \cite{Zurn2} and create 
a multi-well trap by switching on further microtraps \cite{Zurn3}. 

The above-mentioned findings suggest that bosonic systems 
confined in a lattice potential with a superimposed harmonic trap can be used for state preparation in the limit of strong harmonic confinement.
A natural continuation of the present work is to consider time-dependent quench protocols, such as linear quenches or pulsed sequences consisting of abrupt quenches, that may yield 
a substantial improvement on the state preparation, e.g. by exploiting the Landau-Zener mechanism \cite{Landau,Zener,Stueckelberg}.   
Another prospect is to study the case where the parity symmetry is broken by a shift of the harmonic oscillator relative to the lattice. In this case states of opposite parity couple and one 
can induce transitions between states of the zeroth band and states in the first excited band.   
In this context novel kinds of dynamics such as Bloch-like 
oscillations \cite{Dahan,Morsch,Kling} and the cradle mode \cite{Mistakidis,Mistakidis1} can be imprinted to the system.
Finally, another interesting perspective is the study of non-integrability
since the many-body spectrum of the considered system shows a plethora of avoided crossings even in the few-body case.
The latter might be a precursor of the advent of thermalization \cite{Rigol_therm} for larger particle numbers and system sizes.

\section*{Acknowledgments}

The authors thank M. Kurz, G. Z{\"u}rn, C. V. Morfonios, A. V. Zampetaki and H.-D. Meyer for helpful discussions. 
This work has been financially supported by
the excellence cluster ``The Hamburg Centre for Ultrafast Imaging - Structure, Dynamics and Control of Matter at the Atomic Scale'' and 
by the SFB 925 "Light induced dynamics and control of correlated quantum systems" of the Deutsche Forschungsgemeinschaft (DFG). 

\appendix

\section{The Computational Method: MCTDHB} \label{subsec:MCTDH}

Our approach to solve the many-body Schr\"{o}dinger equation $\left( {i\hbar {\partial _t} - H} \right)\Psi (x,t) = 0$ relies on the 
Multi-Configuration Time-Dependent Hartree method for bosons \cite{Alon,Alon1} (MCTDHB). 
MCTDHB has been applied extensively in the literature for the treatment of single species structureless 
bosons (see e.g. \cite{Streltsov1,Alon2,Alon3}). The key idea of MCTDHB lies on the usage of a time-dependent (t-d) and variationally optimized 
many-body basis set, which allows for the optimal truncation of the total Hilbert space. 
The ansatz for the many-body wavefunction is taken as a linear combination of t-d permanents $|\vec{n} (t) \rangle$, with time dependent weights $A_{\vec{n}}(t)$. 
Each t-d permanent is expanded in terms of $M$ t-d variationally optimized single-particle functions (SPFs) 
$\left| \phi_j (t) \right\rangle$. For the numerical implementation the SPFs are expanded within a primitive basis $\lbrace \left| k \right\rangle \rbrace$ of dimension    
$M_{p}$.  
The time-evolution of the $N$-body wavefunction under the effect of the Hamiltonian $\hat{H}$ reduces to the determination of the $A$-vector coefficients and the SPFs, which in turn 
follow the variationally obtained equations of motion \cite{Alon,Alon1}. 
Let us note here that in the limiting case of $M=1$, the method reduces to the t-d Gross-Pitaevski equation, while 
for the case of $M=M_{p}$, the method is equivalent to a full configuration interaction approach to the Schr\" odinger equation within the basis $\lbrace | k \rangle \rbrace$.
\begin{figure}[h]
\includegraphics[width=0.45\textwidth]{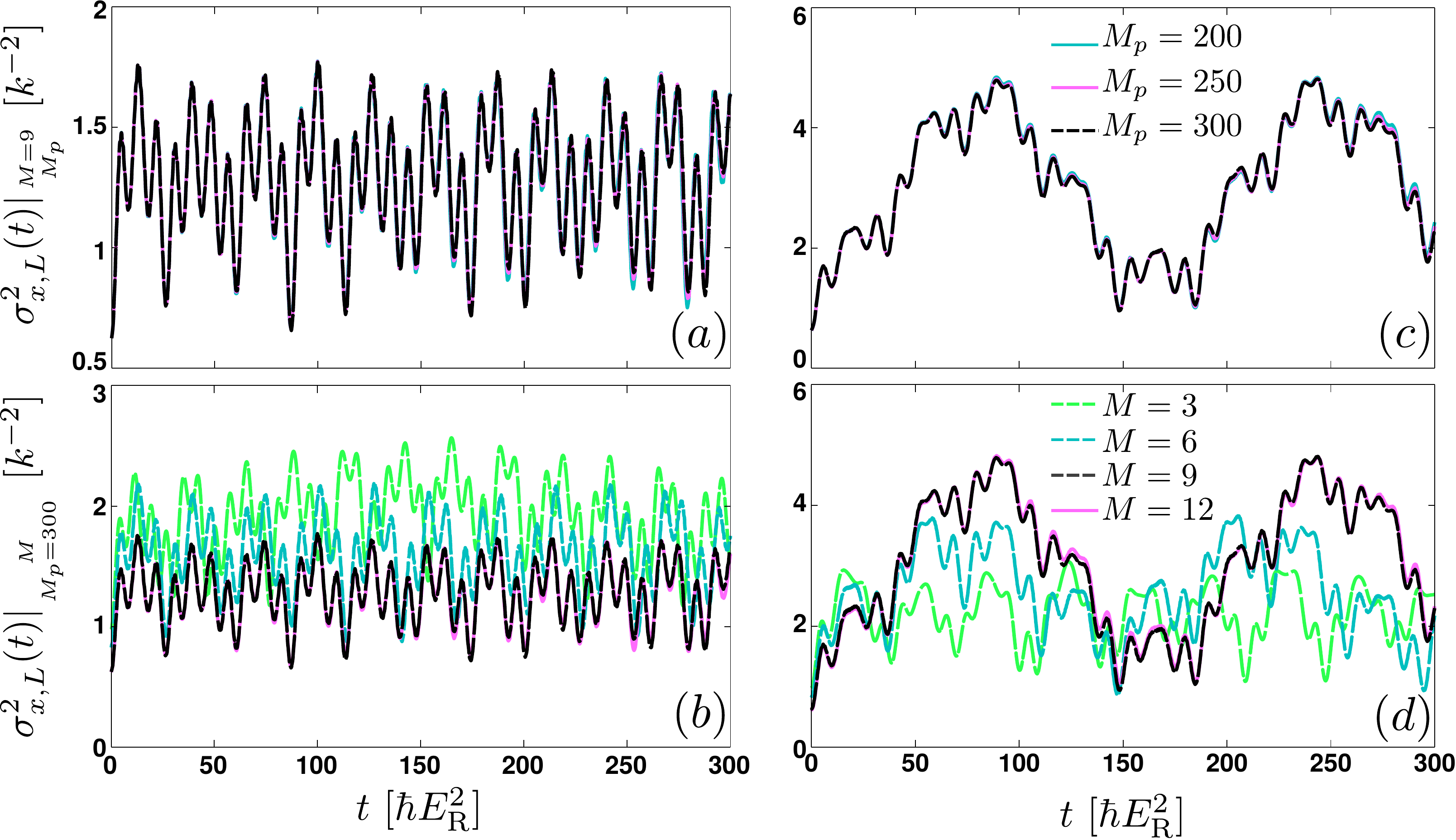}
\caption[]{$\sigma_{x,L}^2(t)$ for a quench from $\omega_i^2= 0.80~ E^2_{\rm R} \hbar^{-2}$ to $\omega^2= 0.40~ E^2_{\rm R} \hbar^{-2}$ and $g=1~E_{\rm R} k^{-1}$ for ($a$) 
different number of grid points $M_p$ and $M=9$ SPFs, ($b$) different number of
SPFs $M$ and $M_p=300$ grid points. 
($c$), ($d$) The same as ($a$), ($b$) but for a quench from $\omega_i^2= 0.80~ E^2_{\rm R} \hbar^{-2}$ to $\omega^2= 0.464~ E^2_{\rm R} \hbar^{-2}$. }
\label{fig:conv}
\end{figure}
For our implementation we have used a sine discrete variable
representation (sin-DVR) as a primitive basis for the SPFs. A sin-DVR intrinsically
introduces hard-wall boundaries at both ends of the potential. 
To obtain the $n$-th many-body eigenstate we rely on the so-called improved relaxation scheme. This scheme can be summarized as follows: a) initialize the system with an ansatz 
set of SPFs $\lbrace |\phi_i^{(0)} \rangle \rbrace$, b) diagonalize the Hamiltonian within a basis spanned by the SPFs,     
c) set the $n$-th obtained eigenvector as the $A^{(0)}$-vector, d) propagate 
the SPFs in imaginary time within a finite time interval $d \tau$, e) update the SPFs to $\lbrace |\phi_i^{(1)} \rangle \rbrace$ and f) repeat steps (b)-(e) until the energy of the state 
converges within the prescribed accuracy.  
To study the dynamics, we propagate the wavefunction by utilizing the appropriate Hamiltonian within the MCTDHB equations of motion.
Finally, let us remark that our implementation has been performed by employing the Multi-Layer Multi-Configuration Hartree method for bosons \cite{Cao,Kronke} (ML-MCTDHB), which reduces 
to MCTDHB for the case of a single bosonic species as considered here.

To verify the numerical convergence of our simulations, we impose the following overlap criteria: 
a) $\left| \langle \Psi | \Psi \rangle - 1 \right| < 10^{-8}$ and b)  
$\left| \langle \varphi_i | \varphi_j \rangle - \delta_{ij} \right| < 10^{-9}$ for the total wavefunction and the SPFs respectively.
Furthermore, we increase the number of SPFs  
and primitive basis states observing a systematic convergence of our
results. For instance, we have used $M_{p}=300$, $M=9$ for the triple-well, $M_{p}=560$, $M=7$ for the seven-well and $M_{p}=600$, $M=6$ for the fifteen-well. 
In the following, we shall briefly demonstrate the convergence behaviour concerning our triple-well simulations either with an increasing number of    
SPFs $M$ (and fixed number of $M_p=300$ grid points) or for a varying number of grid points $M_p$ and a fixed number 
of SPFs, $M=9$. In particular, the fulfillment of the above two conditions is presented below for two different quenches, namely  
from $\omega_i^2= 0.80~ E^2_{\rm R} \hbar^{-2}$ to $\omega^2= 0.40~ E^2_{\rm R} \hbar^{-2}$, $g=1~E_{\rm R} k^{-1}$ and
from $\omega_i^2= 0.80~ E^2_{\rm R} \hbar^{-2}$ to $\omega^2= 0.464~ E^2_{\rm R} \hbar^{-2}$ and $g=1~E_{\rm R} k^{-1}$. 
For reasons of completeness, note that the first of the aforementioned quenches (see Fig. \ref{fig:conv} ($a$), ($b$) lies within the region $II$ (low dynamical response),
and the second (see Fig. \ref{fig:conv} ($c$), ($d$)) in the region $T_{\alpha_3}$ (resonant dynamical response). 
Employing the time-evolution of our main observable, i.e. the position variance $\sigma_{x,L}^2(t)$ (see Eq. \ref{var}), we show (see Fig. \ref{fig:conv} ($a$), ($c$)) that it  
does not alter significantly for varying number of grid points. In particular, even in the case of $M_p=200$ the corresponding differences from the results 
with $M_p=300$ (e.g. presented also in Fig. \ref{fig:3w_dyn_var} ($d$)) are negligible.  
Remarkably enough, the maximum deviation observed in $\sigma_{x,L}^2(t)$, for a quench lying within the region $T_{\alpha_3}$ (see Fig. \ref{fig:conv} ($c$)), calculated using 250 and 300 grid points 
respectively, is of the order of $2.0\%$ for long evolution times ($t>250$).
In addition, we also present for the same quench amplitudes as above, the long time propagation of $\sigma_{x,L}^2(t)$ for different numbers of SPFs (see Fig. 1($b$), ($d$)). 
It is observed that in the case of $M=3$ and $M=6$ strong deviations from the case with $M=9$ occur, while the cases $M=9$ and $M=12$ are 
almost indistinguishable. 
For instance, the maximum deviation observed in $\sigma_{x,L}^2(t)$, for a quench lying in the region $T_{\alpha_3}$ (see Fig. \ref{fig:conv} ($c$)), calculated using 9 and 12 SPFs 
respectively, is of the 
order of $5.0\%$ for long evolution times ($t>200$).
We remark that the same analysis has also been performed for the 
seven- and fifteen-well case (omitted here for brevity) showing a similar behaviour.  
An additional criterion for ensuring the convergence  of our simulations is the population of the lowest occupied SPF, which is kept below $0.01\% $.

{}


\begin{thebibliography}{60}
\bibitem{note} We remark that at the region of the exact crossing (see the dotted box at $\omega^2 \sim 0.21~E_\textrm{R}^2 \hbar^{-2}$ in Fig. \ref{fig:3w_frg}($b$)), 
the $|2,1^{(1)},1\rangle_{S,A}$ dominated eigenstate acquires a contribution from the state $|3,0,1 \rangle_{S,A}$ of the order of $\sim 10 \%$. 
This coupling is caused by the small tilt of the side wells which weakly breaks 
the corresponding parity symmetry. This however does not lead to any considerable  
alterations in the many-body spectrum.
\bibitem{note1} States with two excited particles in the second band do not contribute significantly to the dynamics as they are energetically far from states belonging to the 
ground band.
\bibitem{note2} The dipole mode is mathematically explained by the unitary transformation $| \vec n \rangle (\omega_i) \xrightarrow{U} | \vec n \rangle (\omega)$ which maps the Wannier basis from  
$\omega_i$ to $\omega$ and imprints a coherent-like state to each boson initially residing in a side well.
\end{thebibliography}
\end{document}